\documentclass[useAMS,usenatbib]{mn2e}
\usepackage{color}
\usepackage{graphicx}

\title[The structure of steady, relativistic, magnetised jets with rotation]
{The structure of steady, relativistic, magnetised jets with rotation}
\author[Jos\'e-Mar\'{\i}a Mart\'{\i}]
 {Jos\'e-Mar\'{\i}a Mart\'{\i}$^{1,2}$ \\
  $^1$Departamento de Astronom\'{\i}a y Astrof\'{\i}sica,
  Universitat de Val\`encia, E-46100 Burjassot
  (Val\`encia), SPAIN \\
  $^2$Observatori Astron\`omic, Universitat de Val\`encia,
E-46980 Paterna (Val\`encia), SPAIN}
\date{Released \today}

\pagerange{\pageref{firstpage}--\pageref{lastpage}} \pubyear{2015}

\def\LaTeX{L\kern-.36em\raise.3ex\hbox{a}\kern-.15em
    T\kern-.1667em\lower.7ex\hbox{E}\kern-.125emX}

\begin{document}

\label{firstpage}

\maketitle

\begin{abstract}
  We present equilibrium models of relativistic
magnetised, infinite, axisymmetric jets with rotation propagating
through an homogeneous, unmagnetised ambient medium at rest. The jet
models are characterised by six functions  
defining the radial profiles of density, pressure, and the toroidal
and axial components of velocity and magnetic field. Fixing the
ambient pressure and the jet rest-mass density and axial components of
the flow velocity and magnetic field, we analyze the influence of the 
toroidal magnetic field and several rotation laws on the 
structure of the equilibrium models. Our approach excludes by
construction the analysis of the self-consistently magnetically
launched jet models or the force-free equilibrium
solutions. Several forbidden regions in the magnetic pitch 
angle/magnetization plane are found where models of the class
considered in our study could not be settled. These forbidden regions
are associated with the existence of  maximum axial and toroidal
magnetic field components compatible with the prescribed equilibrium
condition at the jet surface, and/or an excess of centrifugal force
producing gaps with negative pressures in the jet.  The present study can
be easily extended to jet models with different transversal profiles
and magnetic field configurations.

In the last part of the paper, we test the ability of our RMHD code
to maintain steady equilibrium models of axisymmetric RMHD jets in one
and two spatial dimensions. The one dimensional numerical simulations
serve also as a consistency proof of the fidelity of the
analytical steady solutions discussed in the first part of the
paper. The present study allows us to build initial equilibrium jet
models with selected properties for dynamical (and emission)
simulations of magnetised relativistic jets with rotation.
\end{abstract}


\section{Introduction}

  Understanding the physics behind the relativistic jets emanating
from radio-loud AGN \citep{BH12} is a prominent science case that is
pushing continuously the limits of observational capabilities through the
whole electromagnetic spectrum, from radio to gamma-rays. The Event
Horizon Telescope \citep{Cl12}, able to achieve 20 $\mu$as resolution at
submilimeter wavelengths, aimed to capture General Relativistic
signatures near the horizon of the central black hole where these
jets are generated from, is a good example of the continuous improvement in
observational skills. As important as these instrumental/technical
advances are high-resolution numerical simulations that, as a sort of
virtual laboratory, try to connect both theoretical models and more
and more precise observations.

  For parsec scale jets, the strategy of combining analytical steady
jet solutions (acting as initial conditions) with dynamical
and emission simulations has proven already its success in the
relativistic, non-magnetised regime \citep[][Fromm et al. 2015, in
prep.]{GM97,KF97,AG01,AM03,MA09,FP12} 
giving plausible interpretations to most of the phenomenology found in
these objects. 

  Relativistic magnetohydrodynamic simulations have
concentrated on the morphological characterisation of large scale
(i.e., kiloparsec scale) magnetised jets \citep{Ko99b,LA05,KM08,MR10}
and the formation of jets from magnetohydrodynamical mechanisms
\citep[see, e.g.,][]{MB09,TN11,MT12,Po13}. \cite{PF11} computed the 
synchrotron radiation (in the millimeter and submillimeter range) from
RMHD jets through the acceleration region allowing them to address
important issues as the frequency-dependent core-shift effect, or the
signatures of large scale helical fields in the synchrotron
radiation. Previously, \cite{RP08,RP09} presented exploratory results
on the effects of helical magnetic fields on the dynamics of
non-rotating relativistic compact jets and their potential observable
imprints, also using RMHD simulations. \cite{NG10} and \cite{NM14}
interpreted the proper motions of components ejected from the HST-1
complex from radio to optical wavelengths as magnetohydrodynamic shock
fronts making use of one-dimensional RMHD simulations. Without relying
in dynamical simulations, \cite{LP05} considered the polarization
properties of the synchrotron radiation emitted by relativistic,
force-free jets with large scale helical fields and prescribed
emissivity and internal structure, and \cite{BL09} studied the
signatures of relativistic helical motion in the rotation measures of
parsec-scale jets as a way to extract the intrinsic polarization
angles, basing on a very simple helical jet model. Following the same
procedure as in this last work, \cite{BM10} analyzed the
rotation measures associated with jets formed from rapidly
rotating, accreting black holes through the self-consistent general
relativistic MHD simulations of \cite{MB09} extrapolated to parsec scales.

 The main goal of this paper is to build equilibrium models of
relativistic magnetised, axisymmetric jets with rotation propagating
through an homogeneous, unmagnetised ambient medium at rest, which can
serve as initial models for ensuing magnetohydrodynamical simulations
of compact extragalactic jets. Both the jet matter and the ambient
medium are described by an ideal gas equation of state with constant
adiabatic index. As equilibrium solutions, our results complement
those of \cite{ZB08} (for self-similar axisymmetric relativistic,
non-rotating jet models with a purely toroidal magnetic field and an
ultrarelativistic equation of state, in a pressure declining
atmosphere), \cite{GF12} (for non-rotating, relativistic
magnetised jets without surface currents), \cite{BM13} (who 
analyzed the stability of cold relativistic magnetised cylindrical
flows) and \cite{Ly99} (on force-free jets). Beyond
axisymmetric solutions, \cite{ML09,ML12} have investigated 
the influence of jet rotation and shear on the development of the
current-driven kink instability of force-free helical magnetic flows
via three-dimensional relativistic magnetohydrodynamic simulations. 
In the second part of the paper, we use these equilibrium solutions as
initial models for pilot time-dependent RMHD simulations in one and two
spatial dimensions.

  The organization of the paper is as follows. In Sect.~\ref{s:ma} the
basic assumptions of our model are presented. The pressure equation
leading to the transversal equilibrium for both rotating and non-rotating jet models
is derived in Sect.~\ref{s:te}. The remaining jet functions needed to
solve the equilibrium equation are also defined in this section. In
Sect. ~\ref{s:beta-phi-plane}, the resulting equilibrium models are
represented in the averaged magnetic pitch angle/averaged
magnetisation plane and the forbidden regions in this plane
identified. Section~\ref{ss:pm} focusses on the capability of our RMHD
code to maintain steady, one-dimensional solutions along dynamically significant
timescales. In Sect.~\ref{ss:op}, we present the first RMHD simulations of
steady, two-dimensional relativistic, magnetised, rotating jets with
shocks. In Sect.~\ref{s:discussion}, the analytical solutions analyzed in
the first part of the paper and the numerical simulations are put into
an astrophysical context. A brief summary of
the paper along with the most relevant conclusions are presented in  
Sect.~\ref{s:sc}. Finally, a short description of the
  numerical RMHD code used in the simulations and its validation can
  be found in the appendices.

\section{Model assumptions}
\label{s:ma}

  In this paper, we seek solutions of steady, relativistic, magnetised
axisymmetric jets propagating through an unmagnetised ambient medium
at rest. Units are used in which the light speed ($c$), the ambient
density ($\rho_a$) and the radius of the jet ($R_j$) are set to
unity. Both the jet and the ambient medium plasmas are assumed to
behave as a perfect gas with constant adiabatic index $\gamma = 4/3$.

  In order to make the problem tractable, we adopt several
simplifications. As said in the previous paragraph, the jets are
assumed to be axisymmetric. Using cylindrical coordinates
(referred to an orthonormal
cylindrical basis $\{{\bf e}_r,{\bf e}_\phi,{\bf e}_z\}$) in
which the jets propagate along the $z$ axis, axisymmetry implies that
there is no dependence on the azimuthal cylindrical coordinate,
$\phi$. In the first part of the paper, we shall further assume that the
jet models have slab symmetry along the $z$ axis. This means that the
radial magnetic field, $B^r$, is zero. Finally, this symmetry
condition together with the assumed stationarity of the flow, forces
the radial velocity, $v^r$, to be zero too. Hence the jet solutions
are characterised by six functions, namely the density and the
pressure, $\rho(r)$ and $p(r)$, respectively, and the two
remaining components of the velocity, $v^\phi(r)$, $v^z(r)$,
and of the magnetic field, $B^\phi(r)$, $B^z(r)$. The ambient
medium is characterised by a constant pressure, $p_a$ 
(besides $\rho_a = 1$, $v^r_a = v^\phi_a = v^z_a = 0$, $B^r_a =
B^\phi_a = B^z_a = 0$). 

  Under these conditions, the equation of transversal equilibrium
establishing the radial balance between the total pressure gradient, the 
centrifugal force and the magnetic tension, allows to find the
equilibrium profile of one of the variables in terms of the others. We 
shall fix the radial profiles of $\rho$, $v^\phi$, $v^z$, $B^\phi$ and
$B^z$, and solve for the profile of the gas pressure, $p$.
 We use top-hat profiles for $\rho$,
$v^z$ and $B^z$
\begin{equation}
\rho(r) = \left\{ \begin{array}{ll}
\rho_j, & 0 \leq r \leq 1 \\

1,        & r > 1,
               \end{array} \right.
\end{equation}
\begin{equation}
v^z(r) = \left\{ \begin{array}{ll}
v^z_j, & 0 \leq r \leq 1 \\

0,        & r > 1,
               \end{array} \right.
\end{equation}
\begin{equation}
B^z(r) = \left\{ \begin{array}{ll}
B^z_j, & 0 \leq r \leq 1 \\

0,        & r > 1,
               \end{array} \right.
\end{equation}
(where $\rho_j$, $v^z_j$ and $B^z_j$ are constants) and more complex
profiles for the two remaining functions. In particular, for the
azimuthal magnetic field in the laboratory frame we choose
\begin{equation}
B^\phi(r) = \left\{ \begin{array}{ll}
\displaystyle{\frac{2 B_{j, \rm
    m}^\phi (r/R_{B^\phi, \rm m})}{1 + (r/R_{B^\phi, \rm
    m})^{2}}}, & 0 \leq r \leq 1 \\

0,        & r > 1.
               \end{array} \right.
\label{eq:bphi}
\end{equation}
This function represents a toroidal magnetic field that grows
linearly for $r \ll R_{B^\phi, \rm
    m}$, reaches a maximum ($ B_{j, \rm
    m}^\phi$) at $r = R_{B^\phi, \rm
    m}$, then decreases as $1/r$ for $r \gg R_{B^\phi, \rm
    m}$ and is set equal to zero for $r>1$. It is a smooth fit of the
  piecewise profile used by \cite{LP89} \citep[see also][]{Ko99a,LA05}
  and corresponds to a uniform current density for radius $r \ll
  R_{B^\phi, \rm m}$, declining up to $r = 1$, and a return
  current at the jet surface. 

  For the rotation profile we consider three situations:

\begin{enumerate}
\item Models without rotation: $v^\phi(r) = 0$.

\item Models with rigid rotation:
\begin{equation}
v^\phi(r) = \left\{ \begin{array}{ll}
v_{j, \rm m}^\phi r, & 0 \leq r \leq 1 \\

0,        & r > 1.
               \end{array} \right.
\end{equation}
\item Differentially rotating models with a smooth transition between
  an inner jet core with rigid rotation and 
  a Keplerian sheath\footnote{Strictly speaking, the sheath will have a
  Keplerian rotation profile only for $r \gg R_{v^\phi, \rm m}$ (or
  $R_{v^\phi, \rm m} \ll 1$).}:
\begin{equation}
v^\phi(r) = \left\{ \begin{array}{ll}
\displaystyle{\frac{3 v_{j, \rm
    m}^\phi (r/R_{v^\phi, \rm m})}{1 +2 (r/R_{v^\phi, \rm
    m})^{3/2}}}, & 0 \leq r \leq 1 \\

0,        & r > 1.
               \end{array} \right.
\label{eq:DR}
\end{equation}
\end{enumerate}

Note that we associate the rotation of the jet with
  a non-zero azimuthal flow velocity, and not with the rotation of
  the magnetic field lines ($v^\phi_B = v^\phi - v^z B^\phi/B^z$),
  which can be non-zero even in the case of $v^\phi = 0$.

Our assumptions exclude by construction the important
  class of force-free equilibrium solutions considered by, e.g.,
  \cite{Ly99,ML09,ML12}, in which the gas pressure and the matter
  inertia are negligible. In
  the context of the present study, the force-free solutions 
  should be understood as complementary. On the other hand, the
  analysis performed in this paper can be applied to any radial
  profiles of density, and axial and toroidal flow velocity and
  magnetic field, in particular to those derived from the
  self-consistently magnetically launched RMHD models in
  \cite{KB07,KV09} or in \cite{Ly10} papers. However these solutions
  are essentially multidimensional and do not belong to the class of
  solutions discussed in this work.

\section{Transversal equilibrium}
\label{s:te}

  Under the conditions established in the previous Section, the
RMHD equations in (orthonormal) cylindrical coordinates \citep[see, e.g., the Appendix~A
in][]{LA05} reduce to a single ordinary differential equation for the transversal
equilibrium of the jet
\begin{equation}
\label{eq:teq}
  \frac{d p^*}{d r} = \frac{\rho h^* W^2 (v^\phi)^2
  - (b^\phi)^2}{r}.
\end{equation}
In this equation, $p^*$ and $h^*$ stand for the total pressure and the
specific enthalpy including the contribution of the magnetic field
\begin{equation}
\label{eq:p*}
  p^* = p + \frac{b^2}{2}
\end{equation}
\begin{equation}
\label{eq:h*}
  h^* = 1 + \varepsilon + p/\rho + b^2/\rho,
\end{equation}
where $p$ is the fluid pressure, $\rho$ its density and $\varepsilon$ its specific
internal energy. $b^\mu$ ($\mu =
t,r,\phi,z$) are the components of the 4-vector representing the magnetic
field in the fluid rest frame and $b^2$ stands for $b^\mu b_\mu$,
where summation over repeated indices is assumed\footnote{Note
    that we are absorbing the (constant) magnetic permeability, $4
    \pi$, in the definition of the magnetic field.}. $v^i$ ($i = r, \phi,
z$) are the components of the fluid 3-velocity in the laboratory
frame, which are related to the flow Lorentz factor, $W$, according to: 

\begin{equation}
\label{eq:W}
  W = \frac{1}{\sqrt{1-v^i v_i}}.
\end{equation}

The following relations hold between the components of the
magnetic field 4-vector in the comoving frame and the three vector
components $B^i$ measured in the laboratory frame:
\begin{eqnarray}
\label{b0}
  b^0 & = & W B^i v_i \ , \\
  \label{bi}
  b^i & = & \frac{B^i}{W} + b^0 v^i.
\end{eqnarray}
The square of the modulus of the
magnetic field can be written as
\begin{equation}
  b^2 = \frac{{B}^2}{W^2} + (B^i v_i)^2 
\end{equation}
with $B^2 = B^iB_i$.

  Equation (\ref{eq:teq}) establishes the transversal equilibrium
between the total pressure gradient and the centrifugal force (first
term on the r.h.s.), that tends to produce a positive gradient of the
radial total pressure profile, and the magnetic tension (second term
on the r.h.s), that in turn favours to increase the total pressure
towards the axis. Once fixed the radial profiles
of $\rho$, $v^\phi$, $v^z$, $B^\phi$ and $B^z$, we solve the
resulting first-order, linear, non-homogeneous equation for the
profile of the gas pressure, $p$, together with the boundary condition
at $r \! = \! 1$ given by $p^*_1 = p_a$, where $p^*_1$ stands for
$p^*(r \! = \! 1)$\footnote{Along this work, quantities with subindex
  1 refer to their values at $r \! = \! 1$, i.e., $q_1 := q(r \! = \! 1)$.}.

  Unless otherwise established, all the models considered in this paper
have been computed for $\rho_j = 0.01$, $v_j^z = 0.97$, $R_{B^\phi, \rm
  m} = 0.37$, $p_a = 0.1$.

\subsection{Jets without rotation}
\label{ss:te-nrj}

  In the case of a jet without rotation, the
equilibrium equation (\ref{eq:teq}) can be easily transformed into
\begin{equation}
\frac{d p}{d r} = -\frac{(B^\phi)^2}{r
    W^2} - \frac{B^\phi}{W^2} \frac{d B^\phi}{d r},
\end{equation}
which can be integrated by separation of variables to give
\begin{equation}
p(r) = \left\{ \begin{array}{ll}
\displaystyle{2 \left(\frac{B_{j, \rm
    m}^\phi}{W(1 + (r/R_{B^\phi, \rm
    m})^{2})}\right)^2 + C}, & 0 \le r \le 1 \\

p_a, & r > 1.
               \end{array} \right.
\end{equation}
Using the boundary condition, the integration constant $C$ can be
fixed to be
\begin{equation}
C = p_a - \frac{(B_j^z)^2}{2} - \frac{(B^\phi_1)^2}{2W^2}
(1 + (R_{B^\phi, \rm m})^2).
\end{equation}

  It can be easily seen that the presence of a toroidal field like the
one defined in (\ref{eq:bphi}) increases the gas pressure up to $r =
\sqrt{R_{B^\phi, \rm m} ((R_{B^\phi, \rm m})^2 - 
  R_{B^\phi, \rm m} +1)}$ ($\approx 0.53$, for $R_{B^\phi, \rm m} =
0.37$) and decreases it outside so that the averaged gas pressure inside
the jet remains unchanged with respect to the case of zero toroidal
magnetic field (see next Section).

\begin{figure*}
\begin{center}
\begin{minipage}{176mm}
\hspace{0.5cm}
\includegraphics[width=8.8cm,angle=0]{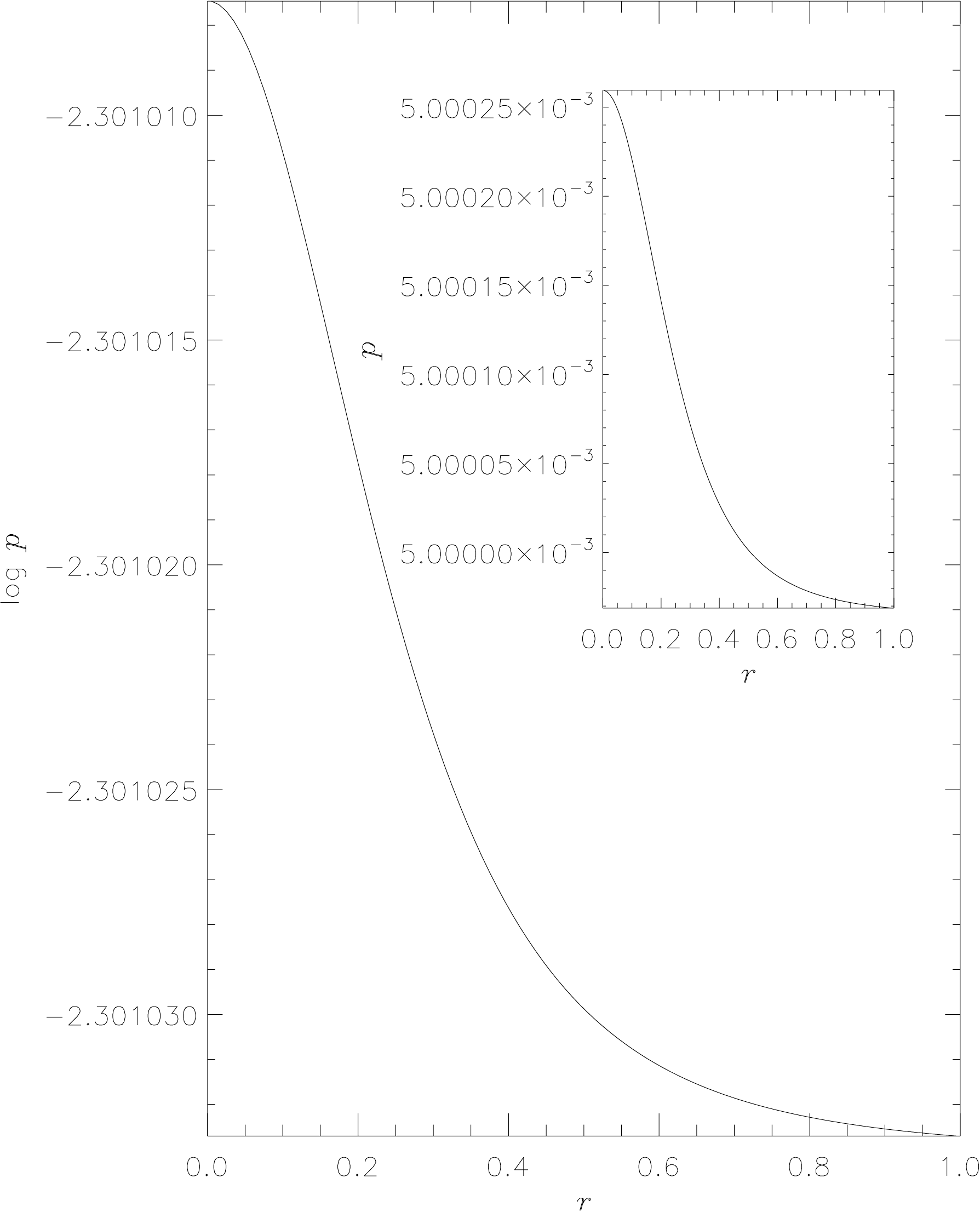}
\includegraphics[width=7.8cm,angle=0]{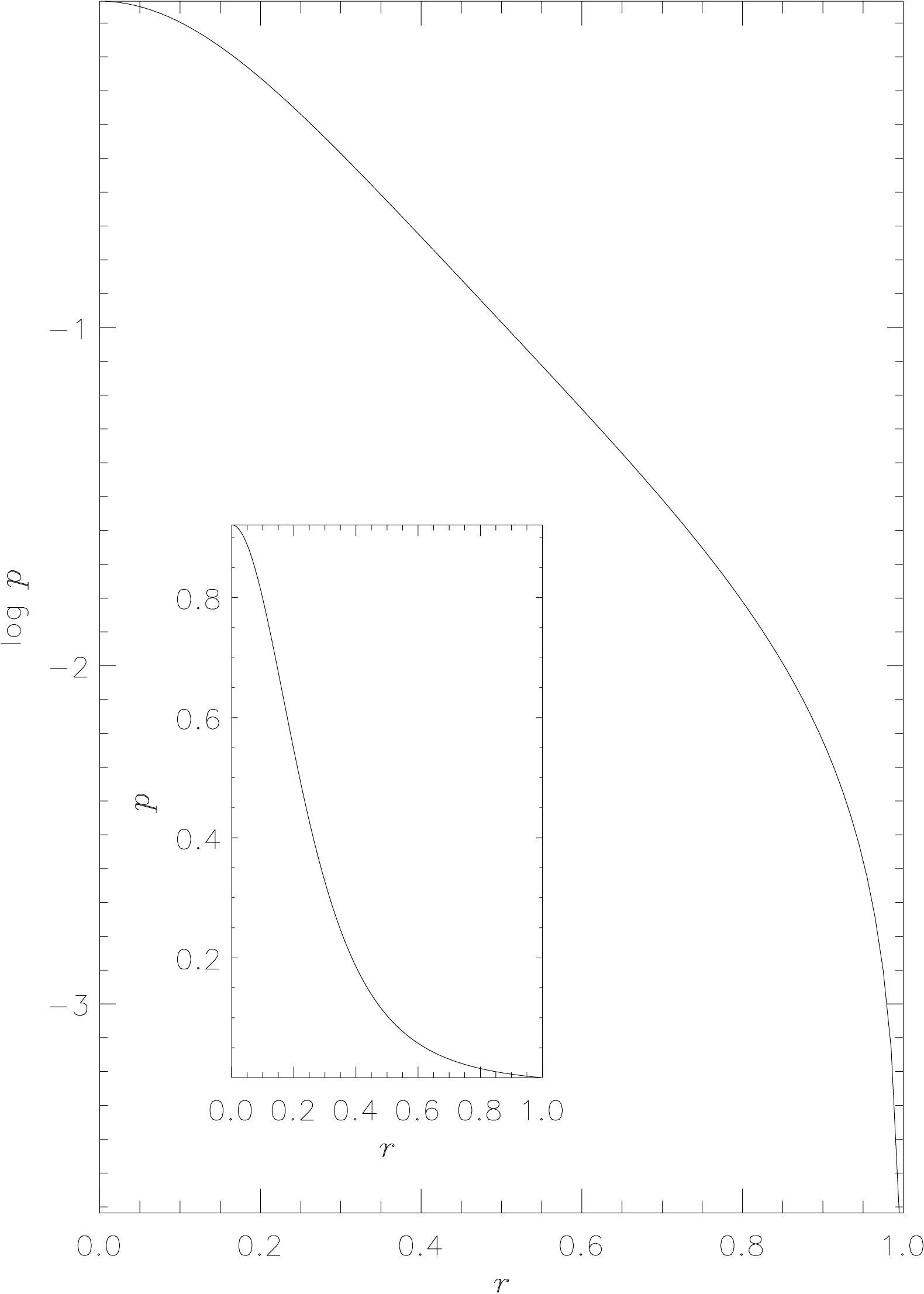}
\caption{Pressure equilibrium profiles for two representative jet models
  without rotation. Model parameters: $\rho_j = 0.01$, 
  $v^z_j = 0.97$, $B^\phi_{j, \rm m} = 1.58 \times 10^{-3}$ (left
  panel), $2.81$ (right panel), $R_{B^\phi, \rm m} =
  0.37$, $B^z_j = 0.436$ (left panel), $4.36 \times 10^{-2}$
  (right panel), $p_a = 0.1$.}
\label{f3}
\end{minipage}
\end{center}
\end{figure*}

  Figure~\ref{f3} displays two representative equilibrium models
of non-rotating jets. In the left panel, the toroidal magnetic field
is small enough as to produce an almost constant gas pressure profile
inside the jet. In the case of the model displayed in the right panel,
the magnetic tension makes the gas pressure drop three orders of
magnitude across the jet\footnote{For the chosen dependence of
  $B^\phi$ with radius, Eq.~(\ref{eq:bphi}), the gradient of the magnetic
  pressure is proportional to (and smaller than) the magnetic tension,
  with the proportionality factor ranging between $(1 + (1/R_{B^\phi,
    \rm m})^2)^{-1}$ ($0.12$ for $R_{B^\phi, \rm m} = 0.37$) and
  $1.0$ across the jet radius.} producing an equilibrium model with a
central spine with high pressure (see inset panel). 

%
%

\subsection{Jets with rotation}
\label{ss:te-rj}

  For jet models with rotation, the equation for the transversal
equilibrium (\ref{eq:teq}) reads
\begin{eqnarray}
\nonumber
\frac{d p}{d r}  & = & -\frac{(B^\phi)^2}{r (W^z)^2} -
\left(\frac{B^\phi}{(W^z)^2} + B^z v^z v^\phi\right)\frac{d B^\phi}{d r} \\
\nonumber
 & & + \left(B^z v^\phi - v^z B^\phi\right) B^z \frac{d  
    v^\phi}{d r} + \frac{\rho h
  W^2 (v^\phi)^2}{r} \\
 & & + \frac{B^z v^\phi}{r} \left(B^z v^\phi - 2 B^\phi v^z\right),
\label{eq:peq_w_r}
\end{eqnarray}
where $(W^z)^2 = \left(1 - (v^z)^2\right)^{-1}$, and $h = 1 + \varepsilon + p/\rho$.
  Equation (\ref{eq:peq_w_r}) is solved numerically, with a standard
fourth-order Runge-Kutta method, starting from the jet surface, where  

%
\begin{equation}
p_1 = p_a - \frac{(B_j^z)^2}{2} \left(1 - (v^\phi_1)^2\right) -
\frac{(B^\phi_1)^2}{2 (W^z)^2}  - v^\phi_1 B^\phi_1 v_j^z B_j^z. 
\end{equation}

\begin{figure*}
\begin{center}
\begin{minipage}{176mm}
\hspace{0.5cm}
\includegraphics[width=7.8cm,angle=0]{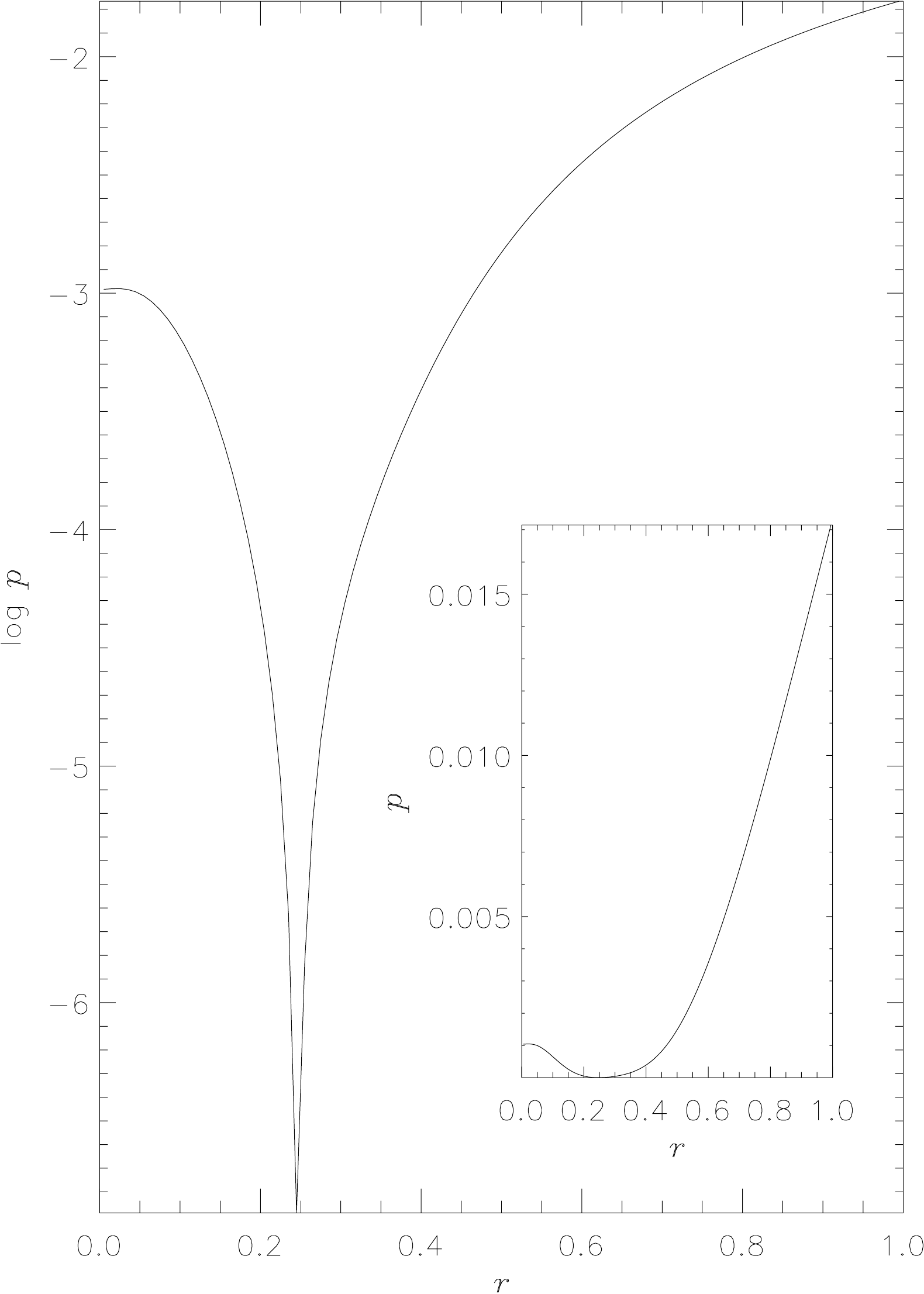}
\includegraphics[width=8.0cm,angle=0]{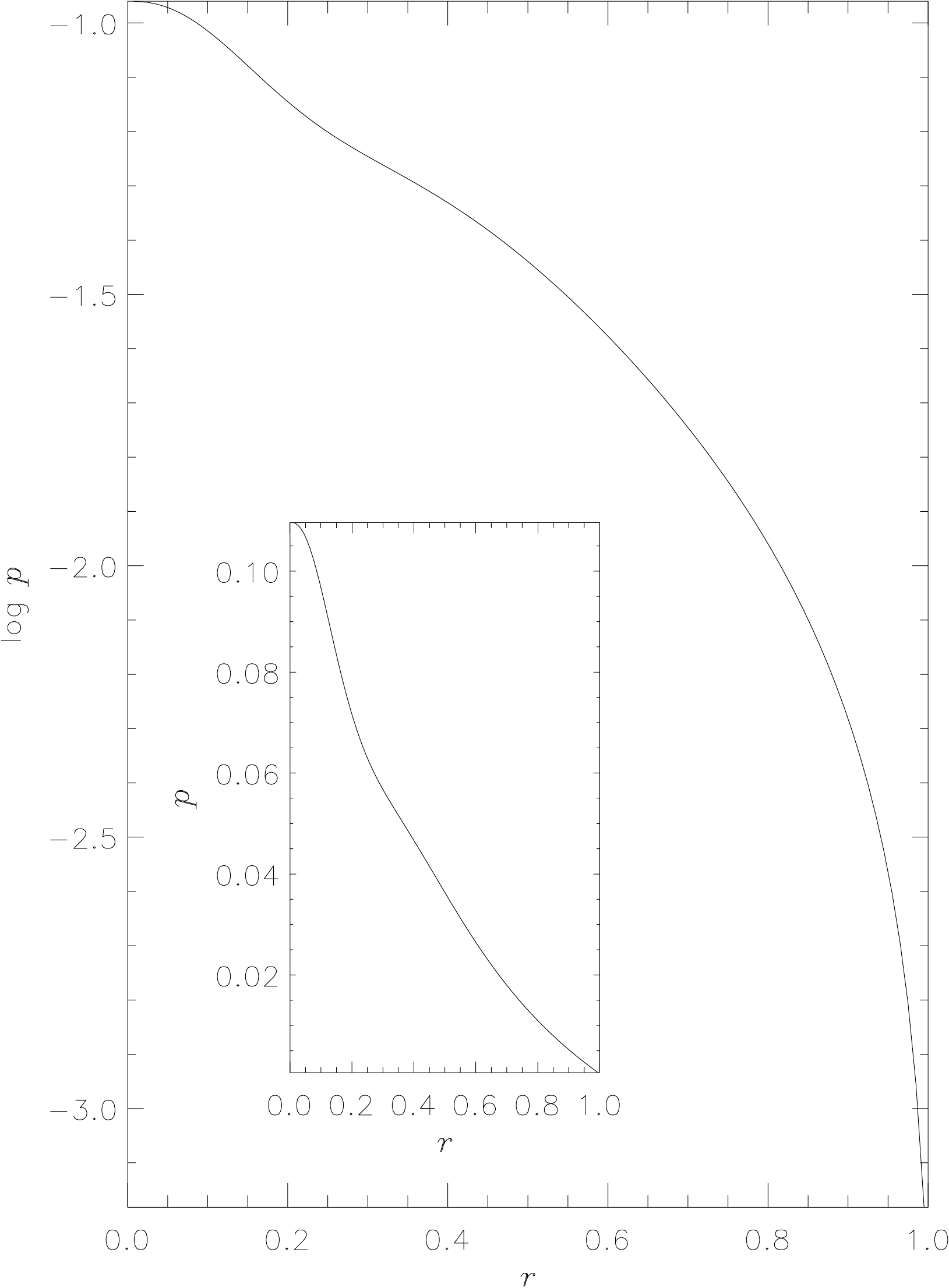}
\caption{Pressure equilibrium profiles for two representative jet models
  with the same rotation profile (rigidly rotating inner
  jet core and a Keplerian sheath). Model parameters: $\rho_j = 0.01$,
  $v^\phi_{j, \rm m} = 0.20$, $R_{v^\phi, \rm m}=0.25$,
  $v^z_j = 0.97$, $B^\phi_{j, \rm m} = 0.151$ (left
  panel), $0.443$ (right panel), $R_{B^\phi, \rm m} =
  0.37$, $B^z_j = 0.396$ (left panel), $4.41 \times 10^{-1}$
  (right panel), $p_a = 0.1$.}
\label{f4}
\end{minipage}
\end{center}
\end{figure*}

  Two illustrative equilibrium models of differentially rotating jets
  ($v_{j, \rm m}^\phi = 0.2$, $R_{v^\phi, \rm m} = 0.25$) are shown in
Fig.~\ref{f4}. In the left panel, the gas pressure profile of the
equilibrium model displays a deep minimum at $r \approx 0.25$ due to the
centrifugal force caused by the jet rotation. Models with larger
azimuthal speeds and/or smaller azimuthal magnetic fields would
make the minimum pressure to reach negative values. In the case
of the model shown in the right panel, the toroidal magnetic field in
the inner region is large enough to keep the pressure high,
despite the action of the centrifugal force.

%
%

\section{Equilibrium models in the $\phi$-$\beta$ plane}
\label{s:beta-phi-plane}

  The equilibrium jet models can be displayed in a magnetic pitch
angle/magnetisation ($\phi-\beta$) plane in terms of the toroidal and axial
magnetic field components, for fixed jet density and kinematics (axial
velocity and rotation velocity profile) and fixed ambient pressure.

  In order to compute the averaged magnetisation and magnetic pitch 
angle across the jet, we first compute the mean values of the
toroidal magnetic field, the rotation velocity and the gas pressure,
namely $\overline{B_j^\phi}$, $\overline{v_j^\phi}$ and
$\overline{p_j}$, according to  
\begin{equation}
\displaystyle{\overline{q_j} := \frac{\int_0^1 q(r) r dr}{\int_0^1 r dr} =
  2 \int_0^1 q(r) r dr}.
\end{equation}

For the averaged toroidal magnetic field, we get
\begin{equation}
\overline{B_j^\phi} = 4 B_{j, \rm m}^\phi R_{B^\phi, \rm
  m} \left(1 - R_{B^\phi, \rm m} \arctan\left(\frac{1}{R_{B^\phi, \rm m}}\right)\right)
\end{equation}
($\approx 0.81 \, B_{j, \rm m}^\phi$, for $R_{B^\phi, \rm m} = 0.37$).

  In the case of models with rigid rotation,
  $\displaystyle{\overline{v_j^\phi} = \frac{2}{3} v_{j, \rm
      m}^\phi}$, whereas for the differentially rotating models,
\begin{equation}
\overline{v_j^\phi} = v_{j, \rm m}^\phi R_{v^\phi, \rm m}^2 \left(
\frac{2}{R_{v^\phi, \rm m}^{3/2}} - \ln\left( \frac{2}{R_{v^\phi, \rm
      m}^{3/2}} + 1 \right) \right)
\end{equation}
($\approx 0.82 \, v_{j, \rm m}^\phi$, for $R_{v^\phi, \rm m} = 0.25$).

  Finally, in the absence of rotation\footnote{Note that
    $\overline{p_j}$ for the non-rotating case is independent of the
    toroidal magnetic field, as advanced in the previous Section.}, 
\begin{equation}
\overline{p_j} = p_a - \frac{(B_j^z)^2}{2}.
\end{equation}
In the rotating cases, the averaged pressure in the jet is computed numerically.

  Now, the averaged pitch angle of the magnetic field, $\overline{\phi_j}$
and the averaged magnetisation, $\overline{\beta_j}$, are defined as
\begin{equation}
\displaystyle{\overline{\phi_j} := \arctan
    \left(\frac{\overline{B_j^\phi}}{B_j^z}\right)},
\end{equation}
\begin{equation}
\displaystyle{\overline{\beta_j} :=
    \frac{\overline{b^2_{j}}}{2 \, \overline{p_j}}}.
\end{equation}
In this last expression, 
\begin{equation}
\overline{b^2_{j}} := \frac{\overline{B_j^\phi}^2}{(W^z)^2} +
  (B_j^z)^2 \left(1 - \overline{v_j^\phi}^2\right) + 2
  \overline{v_j^\phi} \,
\overline{B_j^\phi} v_j^z B_j^z.
\end{equation}

\subsection{Jets without rotation}

\begin{figure*}
\begin{center}
\begin{minipage}{176mm}
\hspace{0.5cm}
\includegraphics[width=8.cm,angle=0]{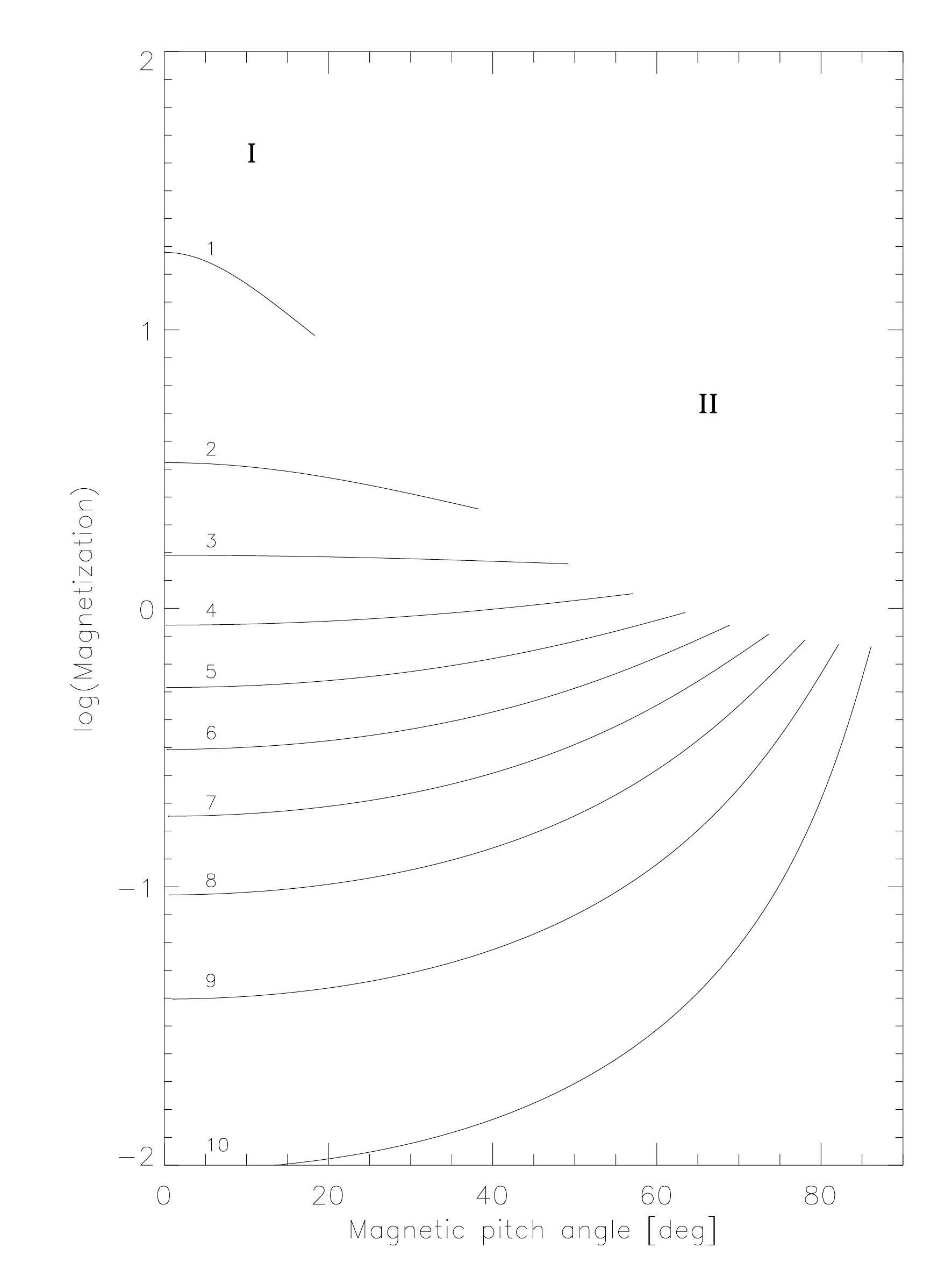}
\includegraphics[width=8.cm,angle=0]{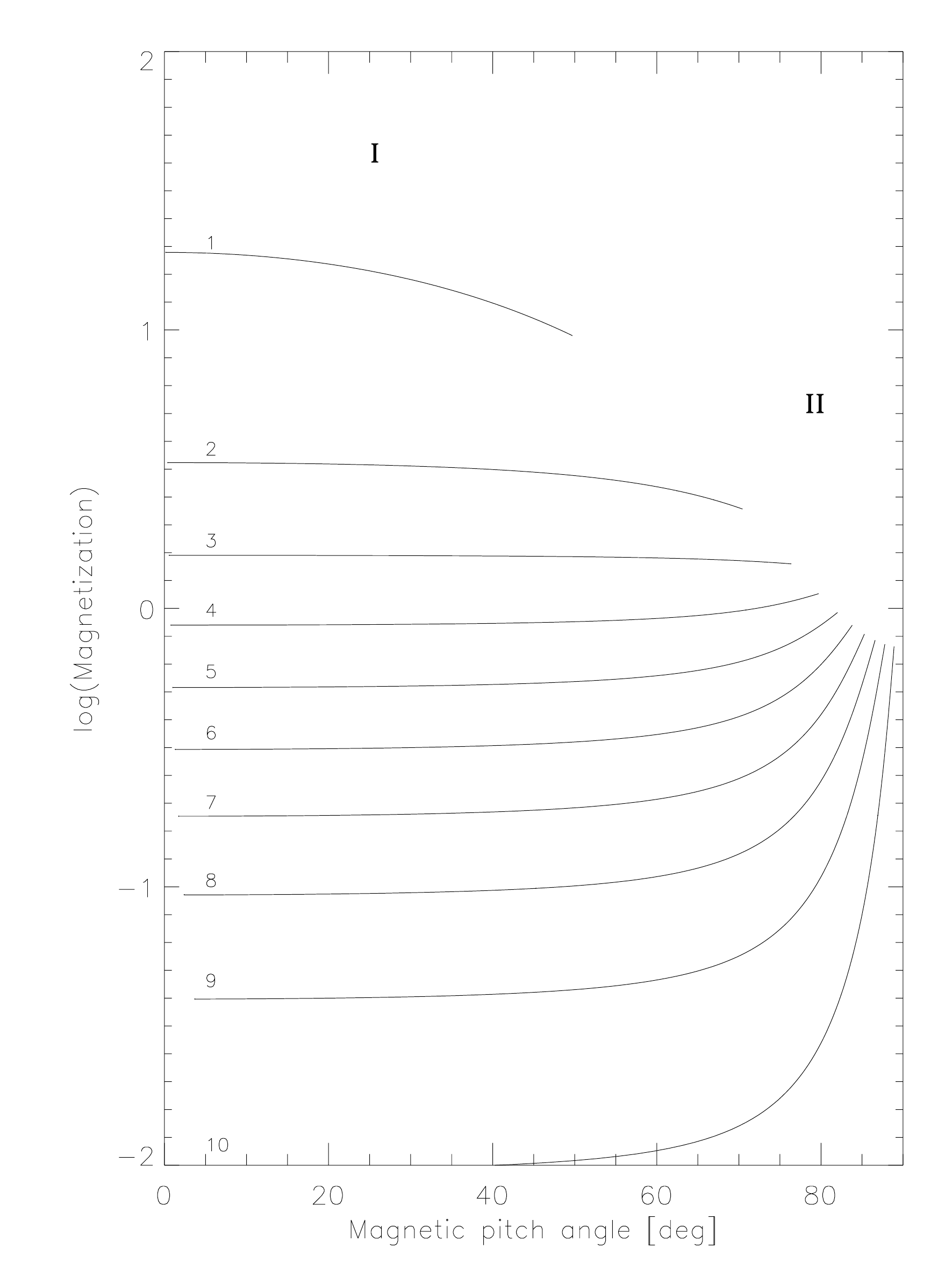}
\caption{Jet magnetisation versus magnetic pitch angle in terms of 
  $B_{j,\rm m}^\phi$ and $B_j^z$ for equilibrium models without
  rotation. Drawn are lines of constant $B_j^z$, with $B_{j,\rm
    m}^\phi$ increasing from left to right along each line up to the
  maximum value compatible with a positive gas pressure at the jet
  surface ($r=1$), for the given axial magnetic field. Model
  parameters: $\rho_j = 0.01$, $v^z_j = 0.50$ (left panel), $0.97$
  (right panel),  $R_{B^\phi, \rm m} = 0.37$, $p_a = 0.1$. Line $i$
  ($=1, \ldots, 10$) corresponds to models with $B_j^z = 0.975 B_{j,\rm
    m}^{z} (11 - i)/10$, where $ B_{j,\rm
    m}^{z} = 0.447$ is the maximum axial magnetic field compatible
  with a positive gas pressure within the jet (and zero toroidal
  magnetic field). Forbidden regions (see text for definitions) are also indicated.}
\label{f1}
\end{minipage}
\end{center}
\end{figure*}

  Figure~\ref{f1} displays non-rotating equilibrium jet models in the
magnetic pitch angle/magnetisation plane for fixed jet density
($\rho_j = 0.01$) and ambient pressure ($p_a = 0.1$), and two axial
jet velocities ($v_j^z = 0.5$, left panel; $v_j^z = 0.97$, right
panel). Drawn are lines of constant $B_j^z$, with $B_{j,\rm m}^\phi$
increasing from left to right along each line (the radius at which the
toroidal magnetic field reaches its maximum, $R_{B^\phi, \rm m}$,
has been fixed to $0.37$ for all the models). There is a maximum axial
magnetic field, $\displaystyle{B_{j,\rm m}^z = \sqrt{2p_a}}$,
corresponding to a purely axial magnetic field compatible with a
positive gas pressure within the jet. The first line starting from the top
corresponds to $B_{j,\rm m}^{z\prime} = \alpha B_{j,\rm m}^z$
($\alpha = 0.975$). Then, the axial magnetic field labeling each line
decreases linearly from top to bottom down to
$\displaystyle{\frac{B_{j,\rm m}^{z\prime}}{10}}$. Along each line,
$B^\phi_1$ increases from left to right up to the largest
value compatible with a positive gas pressure at the jet surface, for
the given $B_j^z$. By construction, the region beyond the line
corresponding to the maximum axial magnetic field can not be filled
with equilibrium jet models within the class of
  solutions sought in this work (forbidden region I). 

  Note that the maximum axial magnetic field, $B_{j,\rm
 m}^z$, defined in the previous paragraph, is independent of the jet
Lorentz factor. Hence, the 
left and right panels of Fig.~\ref{f1} display equilibrium models
covering the same range of axial magnetic fields. On the other hand,
increasing the jet Lorentz factor seems to stretch the lines towards
larger values of the magnetic pitch angle. This is a consequence of
the role played by the Lorentz factor in the magnetic pressure,
reducing the relative weight of the axial magnetic field with respect to
the toroidal one for given magnetisation. Finally, note that the
region beyond the right ends of these $B_{j}^z =$ constant lines
towards the top-right corner of the plot is also forbidden (forbidden
region II), i.e., there are no equilibrium models (within
  our class of solutions) available in this region of the plane (for given
jet density and axial speed, and fixed ambient pressure).

%
%

\subsection{Jets with rotation}

\begin{figure*}
\begin{center}
\begin{minipage}{176mm}
\hspace{0.5cm}
\includegraphics[width=8.cm,angle=0]{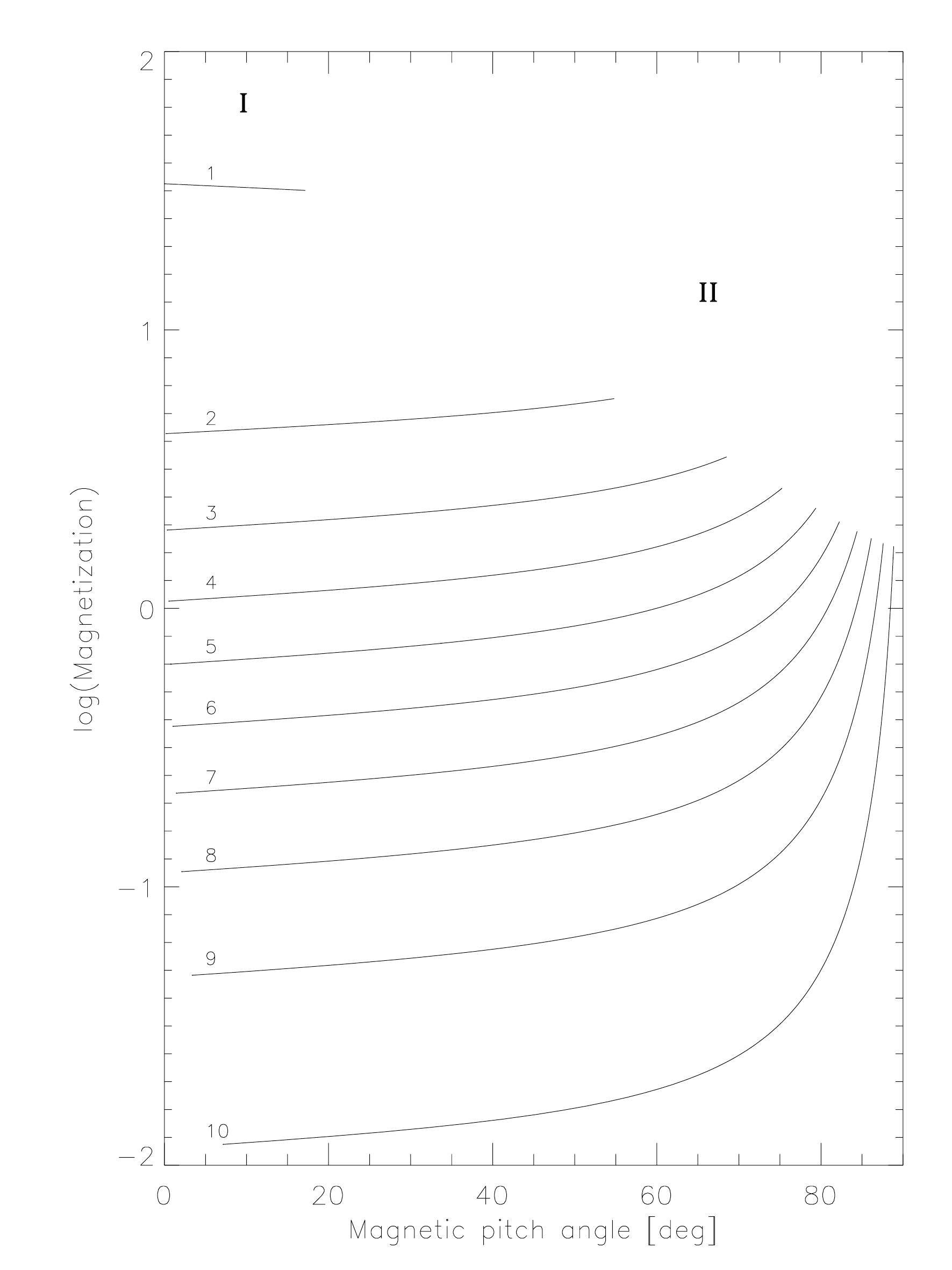} 
\includegraphics[width=8.cm,angle=0]{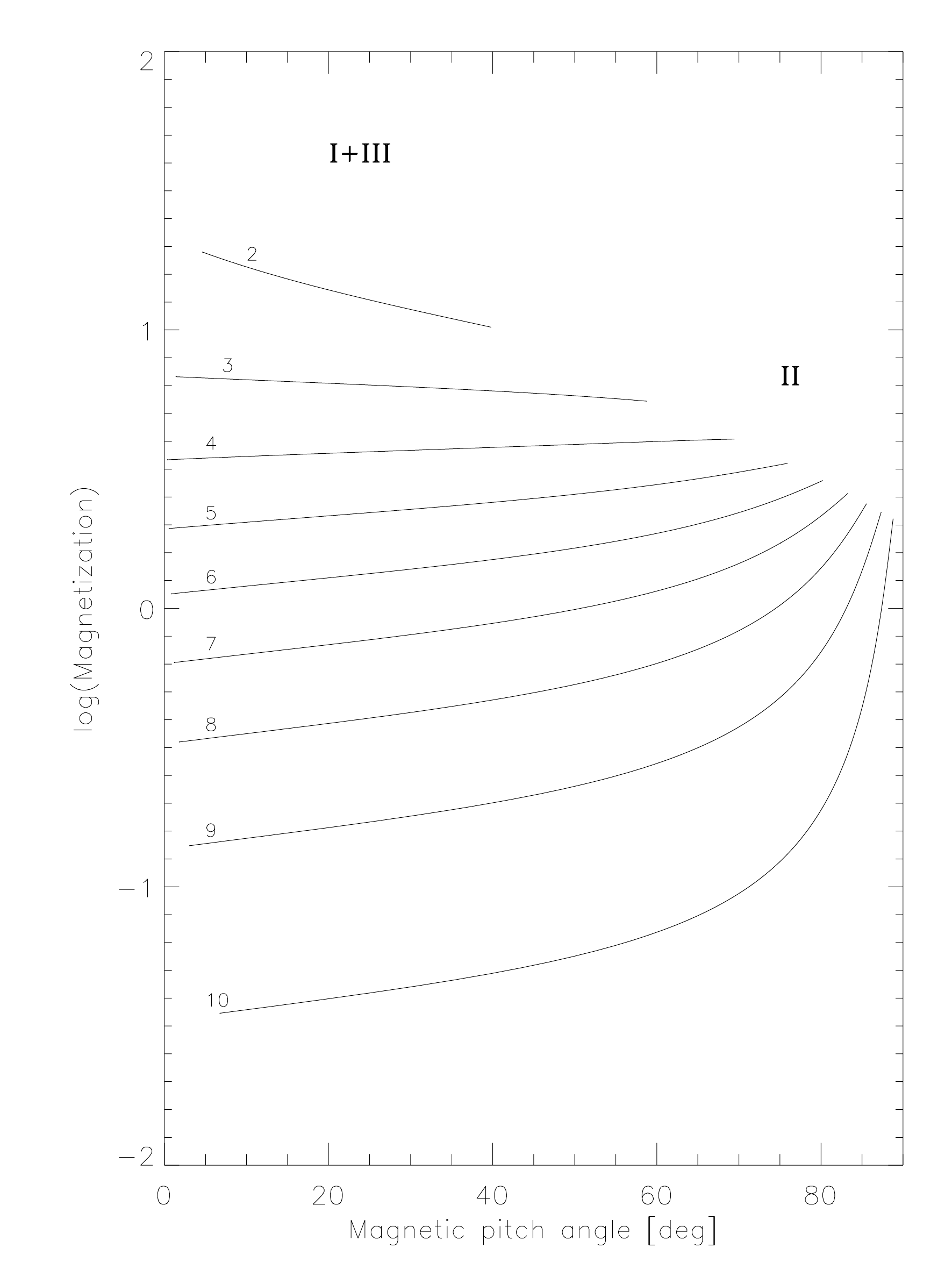} 
\caption{Jet magnetisation versus magnetic pitch angle in terms of 
  $B_{j,\rm m}^\phi$ and $B_j^z$ for equilibrium models with rigid
  rotation in which the helices corresponding to the magnetic field
  and the fluid stream lines turn in the same direction. Drawn are
  lines of constant $B_j^z$, with $B_{j,\rm m}^\phi$ increasing from
  left to right along each line up to the maximum value compatible
  with a positive gas pressure at the jet surface ($r=1$), for the
  given axial magnetic field and rotation speed. Model 
  parameters: $\rho_j = 0.01$, $v^\phi_{j, \rm m} = 0.10$ (left
  panel), $0.20$ (right panel), $v^z_j = 0.97$, $R_{B^\phi, \rm m} =
  0.37$, $p_a = 0.1$. Line $i$ 
  ($=1, \ldots, 10$) corresponds to models with $B_j^z = 0.975 B_{j,\rm
    m}^{z} (11 - i)/10$, where $ B_{j,\rm m}^{z} = 0.449$ (left
  panel), $0.462$ (right panel) is the maximum axial magnetic field
  compatible with a positive gas pressure within the jet for the given
  rotation speed (and zero toroidal magnetic field). Models
  corresponding to line 1 in the right panel have negative gas
  pressures due to an excess of centrifugal force and are not
  shown. Forbidden regions (see text for definitions) are also
  indicated.} 
\label{f2}
\end{minipage}
\end{center}
\end{figure*}

\begin{figure*}
\begin{center}
\begin{minipage}{176mm}
\hspace{0.5cm}
\includegraphics[width=8.cm,angle=0]{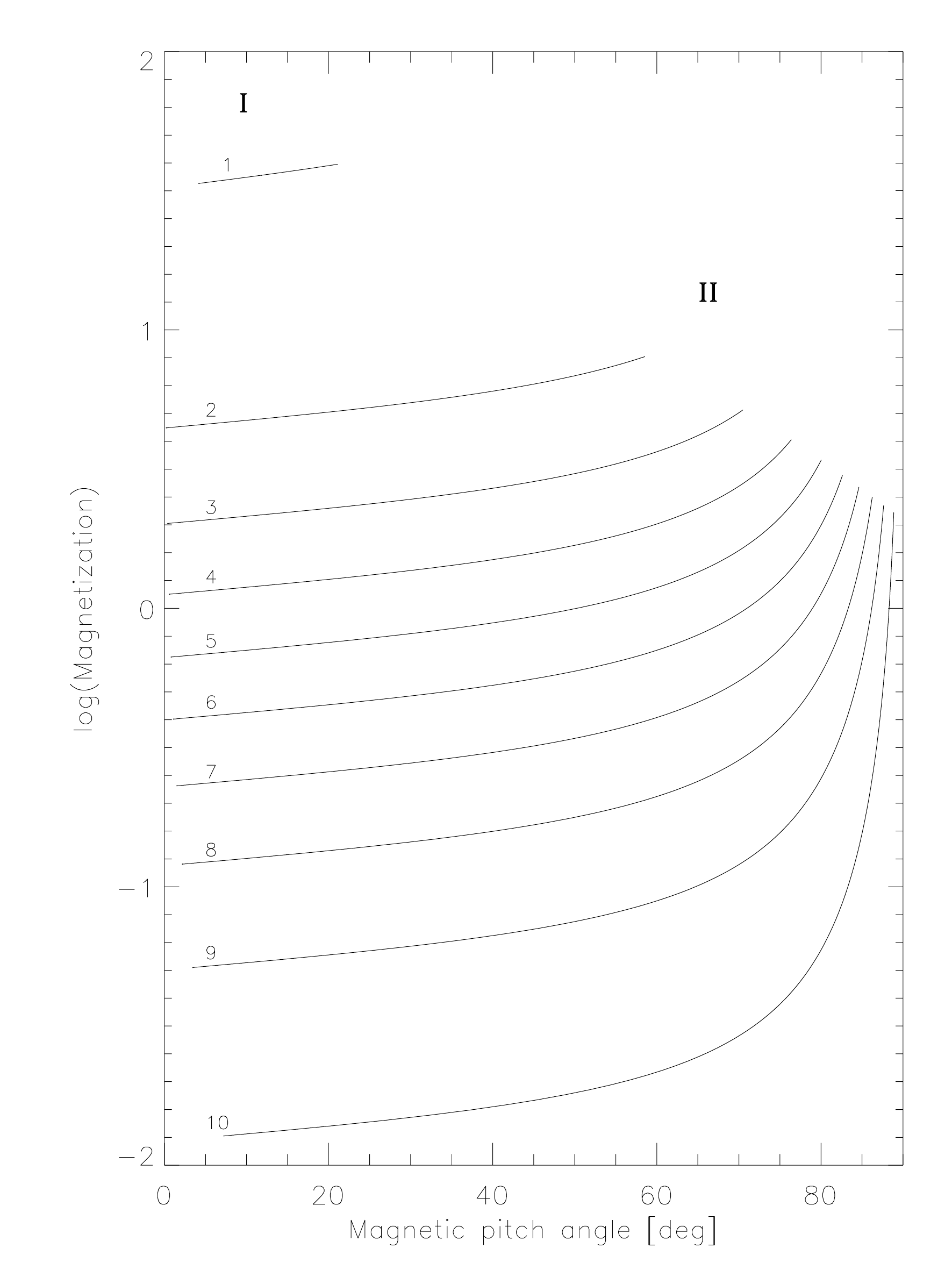}
\includegraphics[width=8.cm,angle=0]{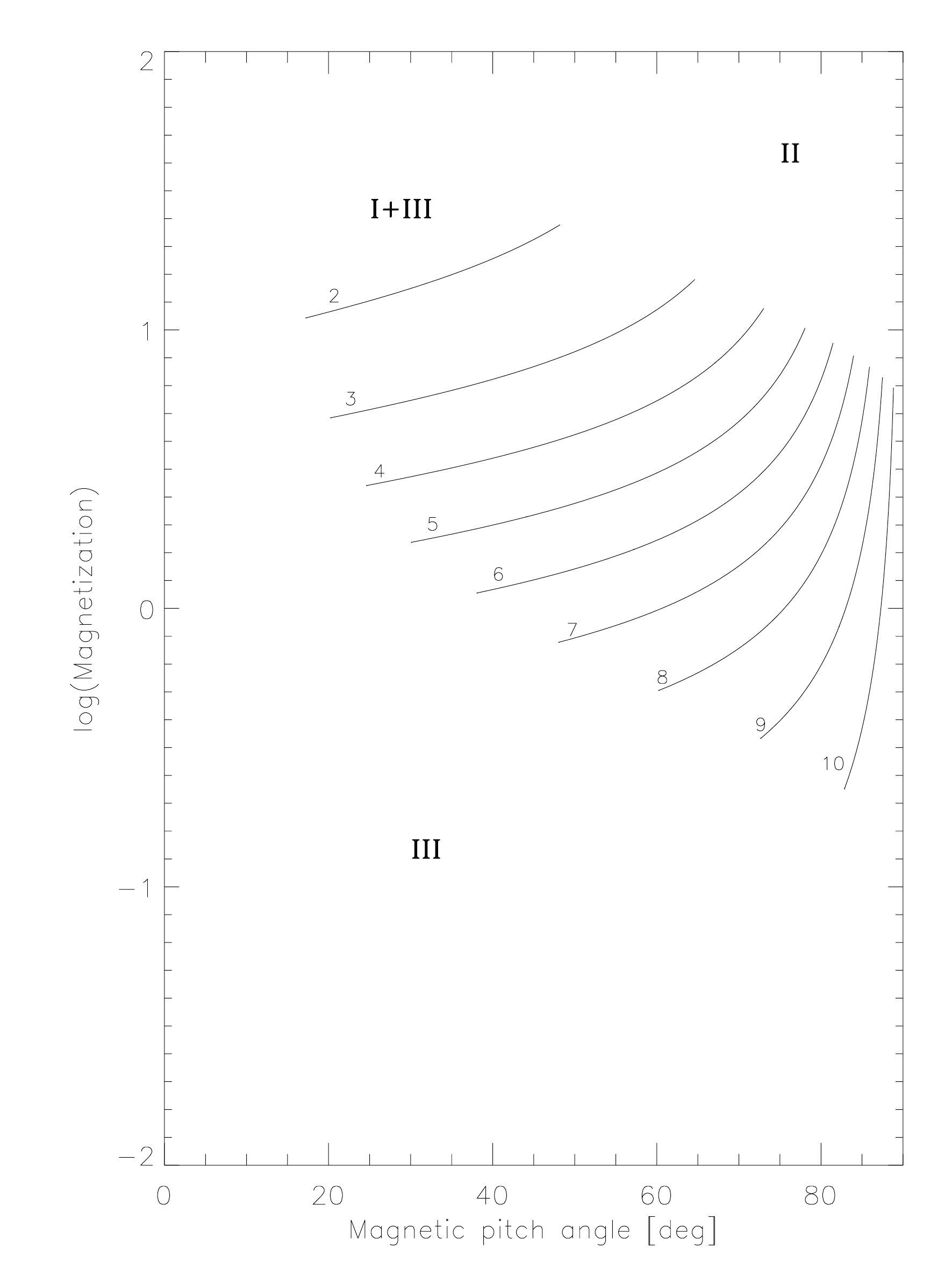}
\caption{Jet magnetisation versus magnetic pitch angle in terms of 
  $B_{j,\rm m}^\phi$ and $B_j^z$ for equilibrium models with
  differential rotation in which the helices corresponding to the magnetic field
  and the fluid stream lines turn in the same direction. Drawn are
  lines of constant $B_j^z$, with $B_{j,\rm m}^\phi$ increasing from
  left to right along each line up to the maximum value compatible
  with a positive gas pressure at the jet surface ($r=1$), for the
  given axial magnetic field and rotation speed. Model 
  parameters: $\rho_j = 0.01$, $v^\phi_{j, \rm m} = 0.10$ (left
  panel), $0.20$ (right panel), $R_{v^\phi,
    \rm m}=0.25$, $v^z_j = 0.97$, $R_{B^\phi, \rm m} = 0.37$, $p_a = 0.1$. Line $i$
  ($=1, \ldots, 10$) corresponds to models with $B_j^z = 0.975 B_{j,\rm
    m}^{z} (11 - i)/10$, where $ B_{j,\rm m}^{z} = 0.448$ (left
  panel), $0.452$ (right panel) is the maximum axial magnetic field
  compatible with a positive gas pressure within the jet for the given
  rotation speed (and zero toroidal magnetic field). Models
  corresponding to line 1 in the right panel have negative gas
  pressures due to an excess of centrifugal force and are not
  shown. Forbidden regions (see text for definitions) are also
  indicated.} 
\label{f2prime}
\end{minipage}
\end{center}
\end{figure*}

  Equilibrium models corresponding to rotating jets are displayed in
Figs.~\ref{f2} and \ref{f2prime}. Figure~\ref{f2} corresponds to models with rigid
rotation whereas Fig.~\ref{f2prime} displays equilibrium models with an
inner jet core with rigid rotation shrouded by a Keplerian sheath
(junction radius, $R_{v^\phi, \rm m}$, equal to $0.25$). The
remaining parameters are as in Fig.~\ref{f1} ($\rho_j = 0.01$, $v_j^z
= 0.97$, $p_a = 0.1$) exception made of the maximum azimuthal
velocity, $v_{j, \rm m}^\phi$, which was set to $0.1$ (left panels in
Figs.~\ref{f2} and \ref{f2prime}) and $0.2$ 
(right panels). Again as in Fig.~\ref{f1}, drawn are lines of constant $B_j^z$,
with $B_{j,\rm m}^\phi$ increasing from left to right along each
line. 

  The maximum axial magnetic field corresponding to a purely axial
magnetic field compatible with a positive gas pressure within the jet is
now
\begin{equation}
\displaystyle{B_{j,\rm m}^z = \sqrt{\frac{2 p_a}{1 -
      (v^\phi_1)^2}}}.
\end{equation}
Again, the first line starting from the top 
corresponds to $B_{j,\rm m}^{z\prime} = \alpha B_{j,\rm m}^z$
($\alpha = 0.975$). Then, the axial magnetic field labeling each line
decreases linearly from top to bottom up to
$\displaystyle{\frac{B_{j,\rm m}^{z\prime}}{10}}$. Along each line,
$B^\phi_1$ increases from left to right up to the largest
value compatible with a positive gas pressure at the jet surface, for
the given $B_j^z$. Let us note that, unlike the case of non-rotating jet
models, there are now two families of equilibrium models depending on
the sign of the product $v^\phi(r) B^\phi(r)$. Figures~\ref{f2}
and \ref{f2prime} show models of the
family with positive sign in which the helices corresponding to
the magnetic field and the fluid stream lines turn in the same
direction. Although similar concerning the transversal
structure of the equilibrium models, both families can have
qualitatively different emission imprints. The forbidden regions I
(for axial magnetic fields beyond the maximum)  and II (beyond
the right ends of the $B_{j}^z =$ constant lines towards the top-right
corner of the plot) are also present in the case of rotating jets.

  Let us now consider the left panels of Figs.~\ref{f2} and \ref{f2prime}. The
differences among the lines drawn in these panels and those 
displayed in the right panel of Fig.~\ref{f1} are small. This is a 
sign that the azimuthal velocity is small and does not affect much the
structure of the models. By contrast, the lines displayed in the right
panels of Figs.~\ref{f2} and specially in those of Fig.\ref{f2prime}, occupy a
different region in the pitch angle/magnetisation diagram. This is
mainly due to the fact that for rotating jet models there is another
forbidden region  beyond the left ends of the lines and towards the
bottom-left corner. In this region (forbidden region
  III), models of the kind analyzed in this work would have sections
  with negative gas pressure for given azimuthal velocity profile due
  to the centrifugal force. The effect is larger in the case of models with the
maximum azimuthal velocity inside the jet for which the centrifugal
force is larger (for the same $v_{j, \rm m}^\phi$), as those shown in the
right panel of Fig.~\ref{f2prime}. For the same reason, the presence
of a non-zero azimuthal velocity reduces effectively the maximum axial
magnetic field compatible with positive gas pressures at the jet surface,
hence modifying the forbidden region I. Models corresponding to line 1
in the right panels of Figs.~\ref{f2} and \ref{f2prime}  have
negative gas pressures due to an excess of centrifugal force. 

%
%

%
%

\section{Steady relativistic magnetised jets}
\label{s:srmj}

  The analytical solutions discussed in the previous Section are
interesting on their own but limited. On one hand, these solutions are 
restricted to infinite planar-symmetric jets in pressure equilibrium
(which, in particular, hampers the development of radial components of
the flow velocity and the magnetic field). On the other hand, although
extragalactic jets are stable to a certain extent, they are far from
being steady. Hence the need to combine these analytical steady
solutions, that could serve as initial conditions, with dynamical (and
emission) simulations. This strategy has demonstrated its success in
the context of purely relativistic jets and can be naturally extended
to RMHD jets (see the Introduction). 

  In Sect.~\ref{ss:pm} the analytical solutions described in
Sect.~\ref{s:beta-phi-plane} are used to probe the capability of our
RMHD code to maintain steady solutions along dynamically significant
timescales. This is a crucial necessary step if one wants to use 
dynamical simulations as a numerical laboratory and keep control on
the parameters of the simulated models. As a first application, in 
Sect.~\ref{ss:op}, the RMHD code will be used to compute the steady
solutions corresponding to overpressured jets. 

  The numerical RMHD code used in these simulations is a conservative,
finite-volume code based on high-resolution shock-capturing
techniques. Its characteristics, as well as the specific algorithms
used are briefly described in Appendix~\ref{a:1}. Appendix~\ref{a:2}
tests the code performance by means of several 1D and 2D standard
problems.

\subsection{Models with pressure equilibrium: Analytical vs. dynamical
  solutions}
\label{ss:pm}

\begin{figure*}
\begin{center}
\begin{minipage}{176mm}
\includegraphics[width=16.6cm,angle=0]{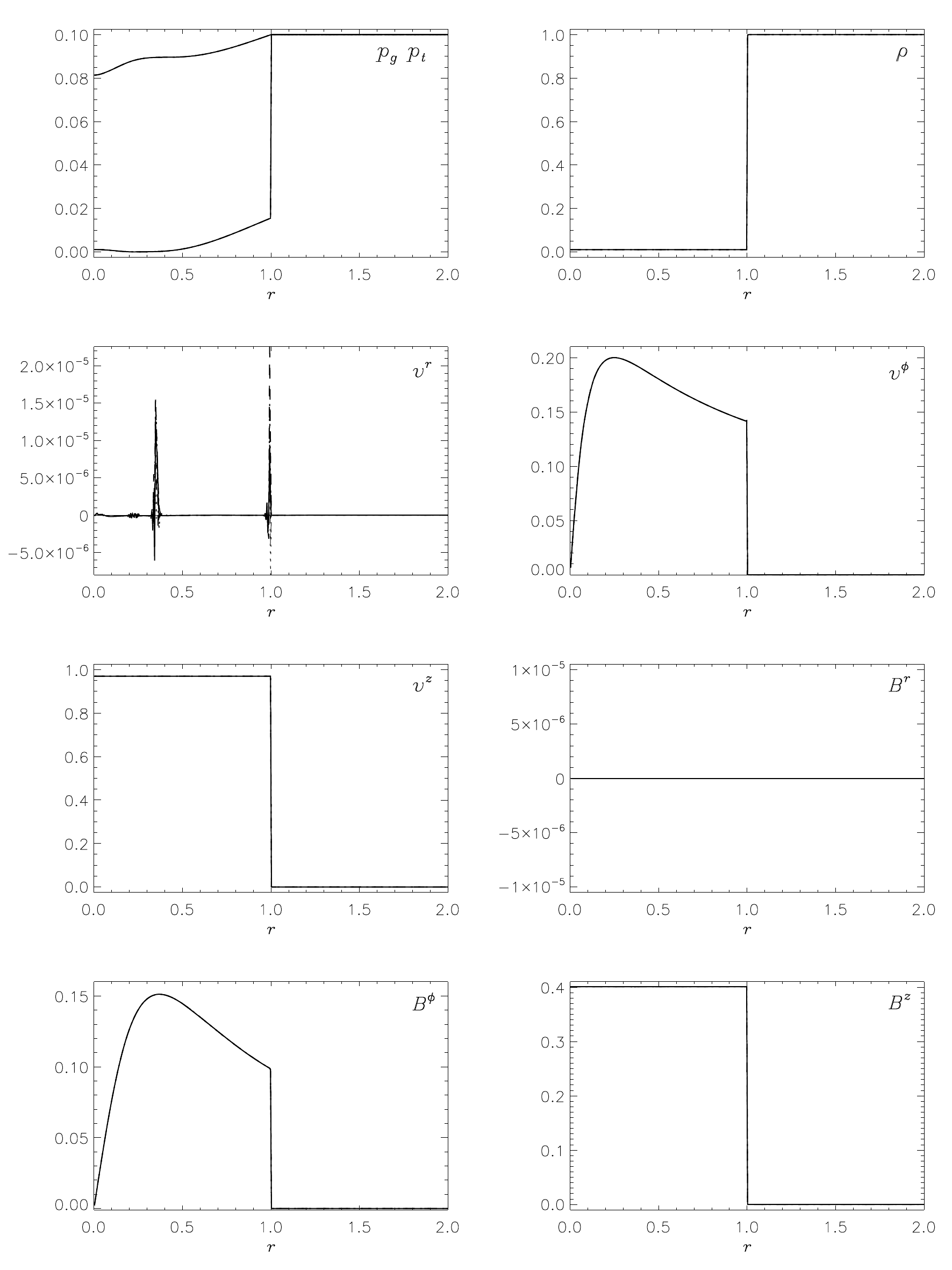}
\caption{Time evolution of the rotating equilibrium model shown in the left
  panel of Fig.~\ref{f4}. Drawn are the initial equilibrium profile 
  (thin filled line) and six lines at $t = 100$ (thin dotted line), $110$
  (thin dashed line), $120$ (thin dotted-dashed line), $130$ (thick filled
  line), $140$ (thick dotted line) and
  $150$ (thick dashed line). The radial computational domain spans
    the interval $r \in [0,4]$. Numerical resolution: 200 cells per jet radius.}
\label{f5}
\end{minipage}
\end{center}
\end{figure*}

\begin{figure*}
\begin{center}
\begin{minipage}{176mm}
\includegraphics[width=16.6cm,angle=0]{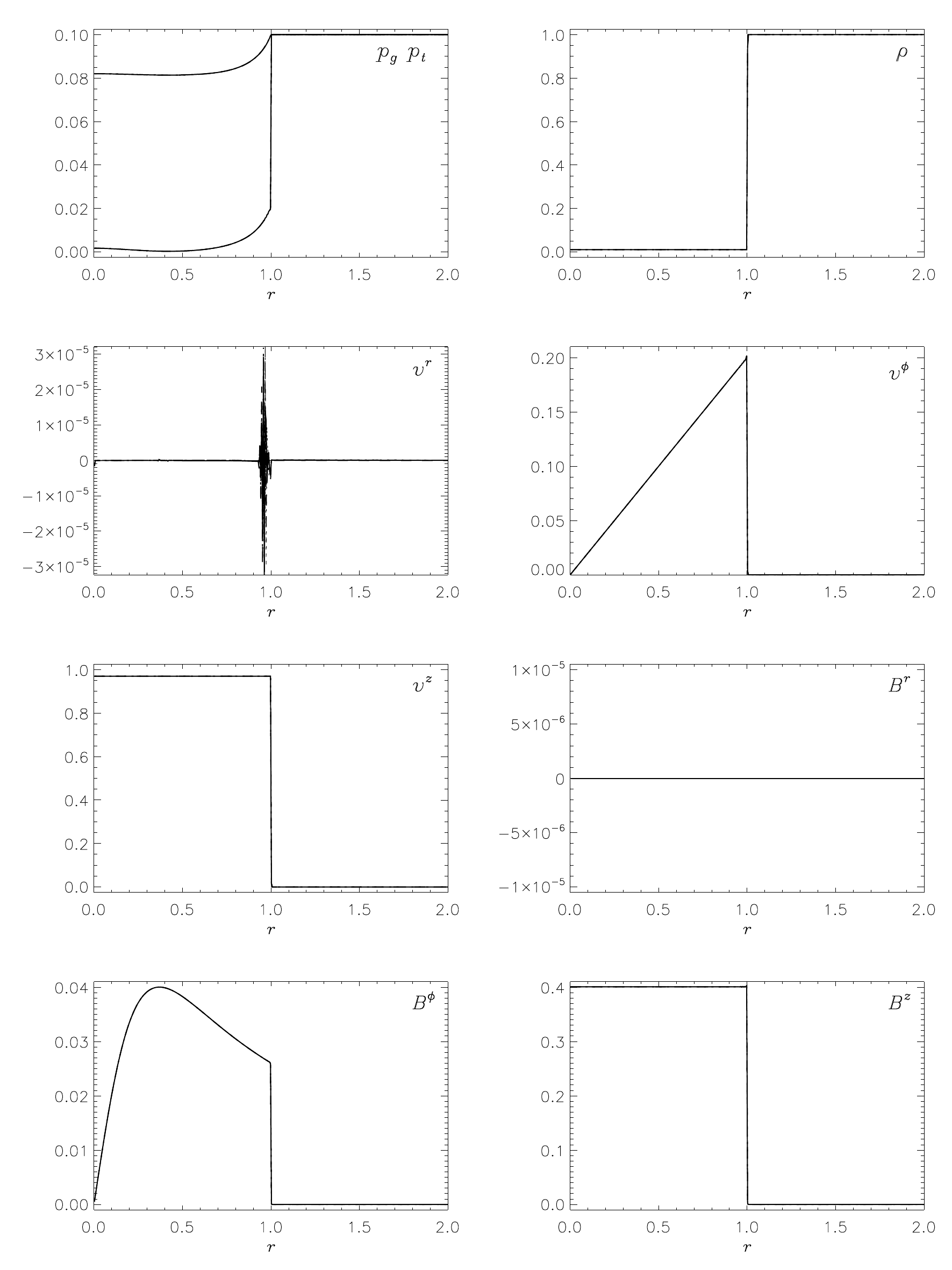}
\caption{Time evolution of an equilibrium model similar to that
  shown in Fig.~\ref{f5} but with rigid rotation (and different
  toroidal magnetic field: $B_{j,\rm m}^\phi = 4.00 \times
  10^{-2}$). Drawn are the initial equilibrium profile  
  (thin filled line) and six lines at $t = 100$ (thin dotted line), $110$
  (thin dashed line), $120$ (thin dotted-dashed line), $130$ (thick filled
  line), $140$ (thick dotted line) and
  $150$ (thick dashed line). The radial computational domain spans
    the interval $r \in [0,4]$. Numerical resolution: 200 cells per jet radius.}
\label{f6}
\end{minipage}
\end{center}
\end{figure*}

  In this case, the jet models have slab symmetry along the $z$ axis
making the problem to find steady solutions one dimensional. We have
used the RMHD code in the 1D radial cylindrical coordinate to test the
code ability to keep steady the initial equilibrium state. Two
revealing cases have been chosen and their evolution shown in
Figs.~\ref{f5} and \ref{f6}. The first case (Fig.~\ref{f5})
corresponds to a model with maximal axial magnetic field, differential
rotation (rigidly rotating inner jet core and Keplerian sheath) and
minimal toroidal magnetic field. The different panels in the figure
display the profiles of gas (and total) pressure and density in
logarithmic scale, and the three components of the fluid velocity and
magnetic field. The initial equilibrium profile (thin filled line) and
six profiles corresponding to times $t = 100, 110, 120, 130,
140, 150$\footnote{Due to relativistic aberration (the jet is flowing
  along  the axis at a speed of $0.97$), the transverse signal
  crossing time of the jet radius is $\approx 4.1$ time units. Hence
  the models shown in the figures correspond to tens of transversal 
  light-crossing times.} are displayed in each of the panels. The code
keeps the initial steady solution (despite there is a small radial
velocity of the order 
of $10^{-5}$ at both the position of the toroidal velocity maximum and
at the jet/ambient medium transition) since the profiles at $t \neq 0$
appear almost 
superimposed. Due to the slab symmetry of the problem (that the code
keeps exactly), the radial component of the magnetic field is
identically zero. The mean relative errors in the jet of the
remaining quantities, ignoring the cell besides the axis (where
the relative errors of the azimuthal components of the velocity and
the magnetic field are $< 3\%$ and $< 5\%$, respectively) and those at
the jet/ambient medium transition (where the errors can be arbitrarily
large), are displayed in Table~\ref{t:t3}. These errors, invisible at
the plot scale, can be reduced by increasing the numerical resolution
(200 numerical cells per jet radius in this case). The difference
between the total energy per unit jet length in the initial analytical
equilibrium model and the model at $t=150$ is of $1.43\%$. 

%
\begin{table*}
 \centering
 \begin{minipage}{172mm}
  \caption{Mean relative errors in the jet for the
    equilibrium models displayed in Figs.~\ref{f5} (differentially
    rotating model, DR) and \ref{f6} (rigidly rotating model, RR) at
    $t = 150$.}
  \begin{tabular}{lrrrrrr}
  \hline
Model & $p$ & $\rho$ & $v^\phi$ & $v^z$ & $B^\phi$ & $B^z$ \\
 \hline
DR & $< 10^{-3}$ & $< 2 \times 10^{-4}$ & $< 4 \times 10^{-6}$ & $< 5
\times 10^{-7}$ & $< 10^{-5}$ & $< 10^{-5}$ \\ 
RR & $< 3 \times 10^{-4}$ & $< 2 \times 10^{-4}$ & $< 10^{-4}$ & $<
10^{-6}$ & $< 5 \times 10^{-5}$ & $< 2 \times 10^{-4}$ \\ 
\hline
\end{tabular}
\label{t:t3}
\end{minipage}
\end{table*}
%

  The model shown in Fig.~\ref{f6} is similar to that
shown in Fig.~\ref{f5} with the same (maximal) axial magnetic field
but with rigid rotation and different (minimal) toroidal magnetic
field. The panel distribution and the lines plotted are the
same as in Fig.~\ref{f5}. The maximum of the azimuthal velocity is
now at the jet surface and the minimum in pressure is shifted towards 
larger radius. The radial velocity is again of the order
of $10^{-5}$ at large times proving that the code has found a steady
solution. The mean relative errors in the jet of the
  remaining quantities (the radial component of the magnetic field is
  zero), excluding the cell besides the axis (where
the errors of the azimuthal components of the velocity and
the magnetic field are $< 7\%$ and $< 8\%$, respectively) and those  at the
  jet/ambient medium transition, are displayed in Table~\ref{t:t3}.
The difference between the total energy per unit jet length in the
initial analytical equilibrium model and the numerical one at $t=150$
is now of $1.18\%$.

%
%

%
%

\subsection{Overpressured jet models: conical shocks and Mach discs}
\label{ss:op}

  In this Section, the equilibrium profiles discussed in
Sect.~\ref{s:beta-phi-plane} are used as a boundary condition to
inject the jets into a two dimensional domain representing an ambient 
medium with a pressure mismatch. In their attempt to reach again the  
equilibrium, the jets undergo sideways motions generating radial 
components of the flow velocity and the magnetic field that break the 
slab symmetry of the original jet model along the $z$ axis. 

%
\begin{table*}
 \centering
 \begin{minipage}{172mm}
  \caption{Parameters defining the overpressured models and
    corresponding averaged jet values.}
  \begin{tabular}{lrrrlccccccccccc}
  \hline
    &  &  & & Rotation  & & & & & & & & & & $\overline{\phi_j}$
& \\
Model & $p_a$ & $K$ & $\rho_j$ & law\footnote{DR: differential
  rotation as defined in Eq.~(\ref{eq:DR}); RR: Rigid rotation.} & $R_{v^\phi, \rm m}$
   & $v_{j, \rm m}^\phi$ & $v_j^z$ & $R_{B^\phi, \rm m}$ & $B_{j, \rm
     m}^\phi$ & $B_j^z$ & $\overline{v_j^\phi}$ &
   $\overline{B_j^\phi}$ & $\overline{p_j}$ & [deg] & $\overline{\beta_j}$\\
 \hline
JO12 & 0.2 & 2.5 & 0.01 & RR &  & 0.15 & 0.97 & 0.37 & 0.5 & 0.5 &
0.100 & 0.407 & 0.246 & 39.14 & 0.602 \\ 
JO13 & 0.2 & 2.5 & 0.01 & RR &  & 0.24 & 0.97 & 0.37 & 0.5 &
0.5 & 0.160 & 0.407 & 0.033 & 39.14 & 4.83 \\ 
\hline
JO85 & 0.2 & 5.0 & 0.01 & DR & 0.25 & 0.34 & 0.92 & 0.37 & 0.5 & 0.5 & 0.280
& 0.407 & 0.358 & 39.14 & 0.503 \\
JO86 & 0.2 & 5.0 & 0.01 & DR & 0.25 & 0.15 & 0.92 & 0.37 & 0.5 & 0.5 &
0.123 & 0.407 & 0.741 & 39.14 & 0.214 \\
\hline
\end{tabular}
\label{t:t1}
\end{minipage}
\end{table*}
%

  Four simulations of overpressured, rotating jets are presented in
this Section. The jets are injected through a nozzle of radius equal
to $1$ into an axisymmetric cylindrical domain with $(r,z) \in [0,4] \times
[0,30]$. The evolution of the flow in the domain is simulated with the
RMHD code (see the Appendix) in 2D radial, axial cylindrical
coordinates with a resolution of 80 (40) cells per jet radius in the
radial (axial) direction. In order to disturb the ambient medium
  as little as possible along the simulation, the domain $(r,z) \in 
  [0,1] \times [0,30]$ is initially filled with the analytical,
  injection solution. Reflecting boundary conditions are set along  
the axis ($r=0$) and at the jet base outside the injection nozzle
($r>1$, $z=0$). Zero gradient conditions are set in the remaining
boundaries. 

\begin{figure*}
\begin{center}
\begin{minipage}{172mm}
\includegraphics[width=16.cm,angle=0]{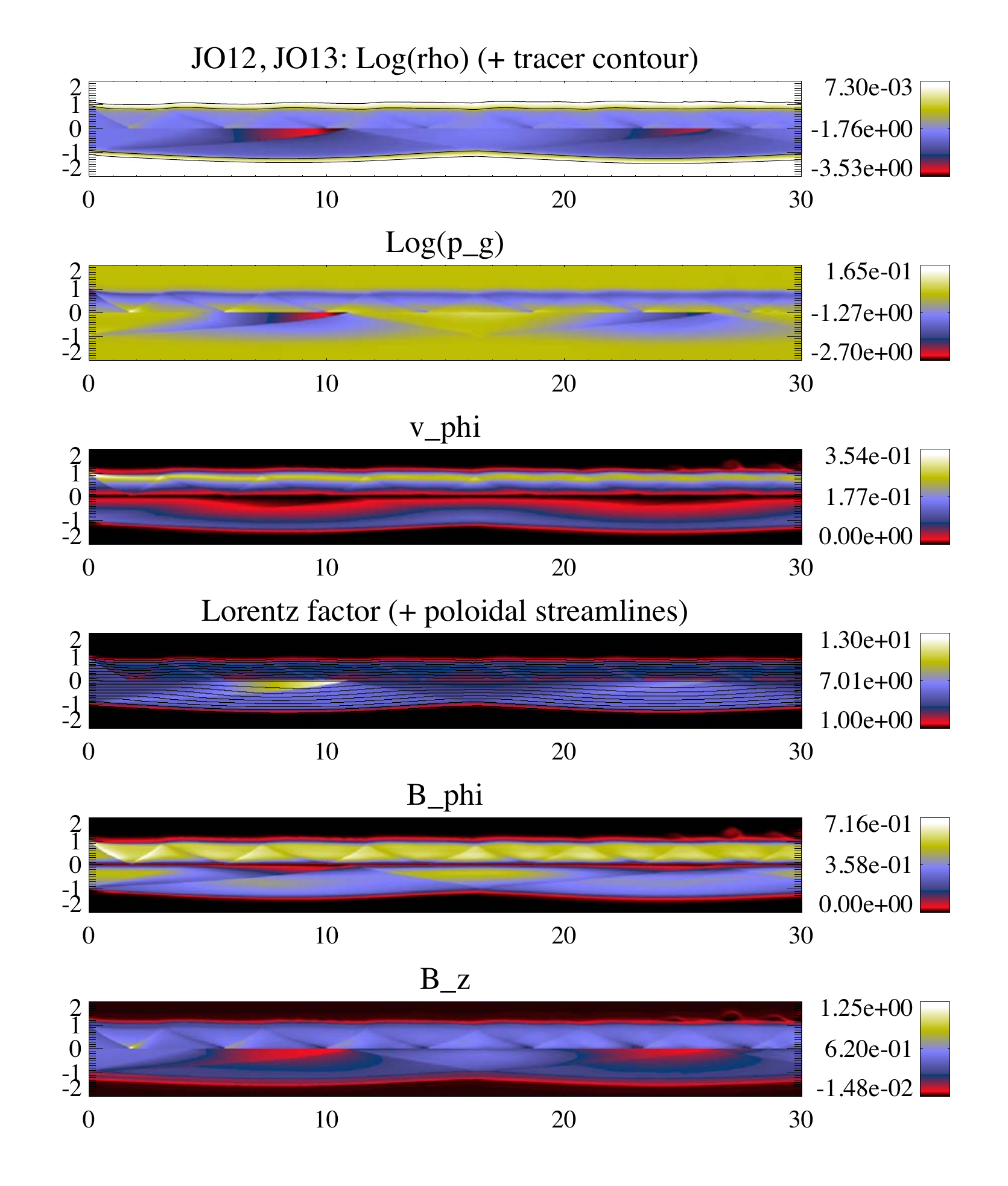}
\caption{Bottom halfs of panels: Model JO12. Top halfs of panels:
  Model JO13. Models parameters: see Table~\ref{t:t1}.  The two
  contours in the top and bottom halfs of the density map establish
  the transition between the jet (inner contour: 90\% of jet material)
  and the ambient medium (outer contour: 10\% of jet material). The steady
  states are set through a series of conical fast-magnetosonic shocks
  (with a spacing of about 15 initial jet radii in model JO12, and of
  about 4 initial jet radii in model JO13).} 
\label{f:JO12-13}
\end{minipage}
\end{center}
\end{figure*}

  The basic parameters defining the models together with
the corresponding averaged jet values are displayed in
Table~\ref{t:t1}.  The models, set up to be in equilibrium with an
ambient pressure $p_a'$, are injected into an atmosphere with pressure 
$p_a = p_a'/K$, where $K$ is the jet overpressure factor. Models JO12
and JO13 are injected with a moderate overpresure factor of $K = 2.5$
and rotate rigidly with maximum speed $v_{j, \rm   m}^\phi = 0.15$
(Model JO12) and $0.24$ (JO13). The rest of the parameters are the
same in the two models. However, the different rotation profiles at
injection generate different equilibrium profiles of the total
pressure which are on the basis of the different flow structure of
both models. Figure~\ref{f:JO12-13} shows the distributions of several
quantities at time $t = 100$. In both cases the new equilibrium states
are set through a series of conical fast-magnetosonic shocks however,
whereas the separation between shocks in model JO12 is about 15
initial jet radii, in the case of model JO13 the separation is much
shorter (of only 4 initial jet radii). Related to this dissimilarity is
also the change in the jet radius, which in the case of model JO13
remains very close to 1, but in the case of JO12 oscillates between 1
and 1.5. This different structure can also point to differences in their
stability properties in response to small perturbations. 

\begin{figure*}
\begin{center}
\begin{minipage}{172mm}
\vspace{-4cm}
\hspace{30mm}
\includegraphics[width=20.6cm,angle=0]{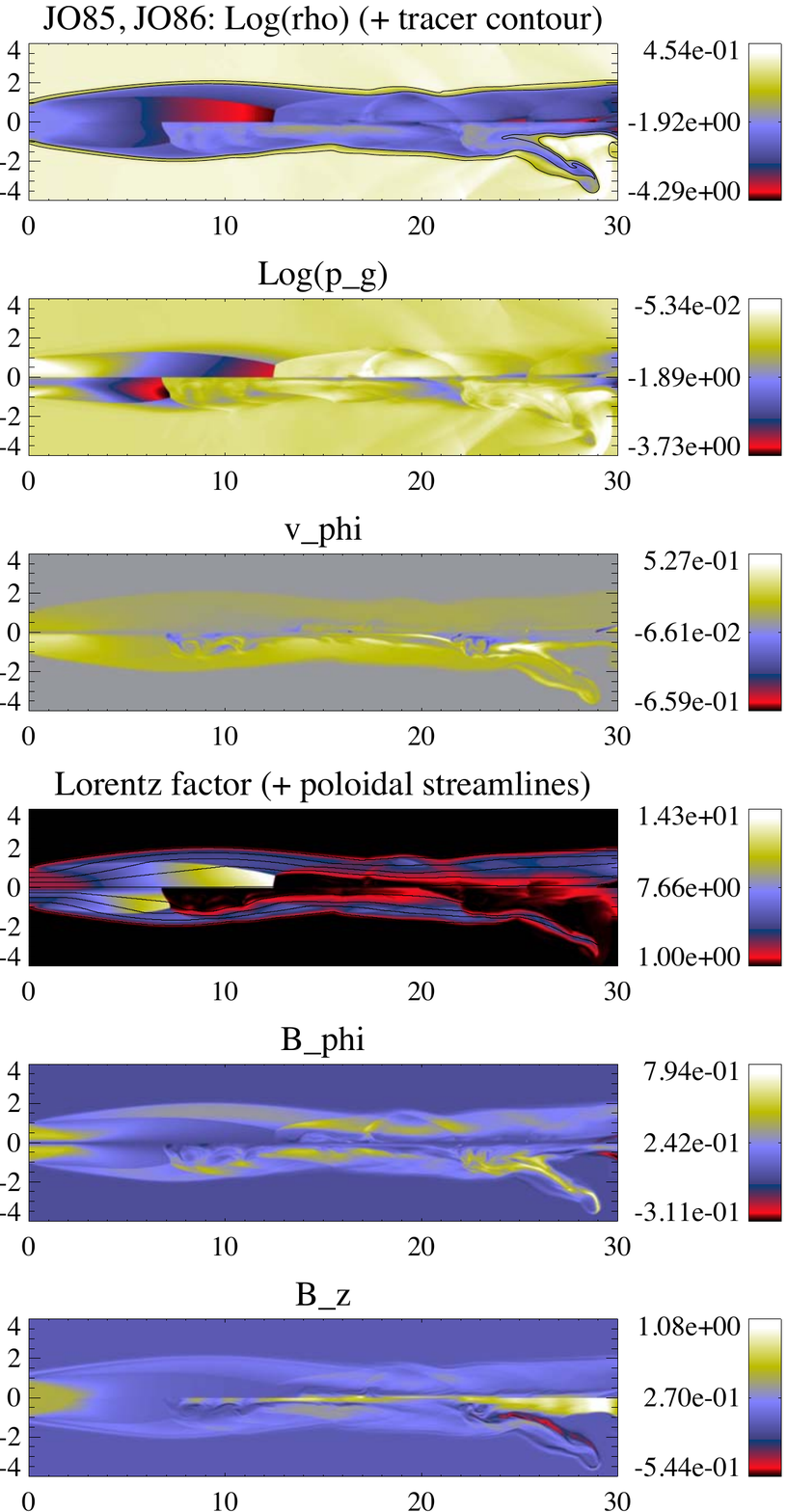}
\caption{Bottom halfs of panels: Model JO85. Top halfs of panels:
  Model JO86. Models parameters: see Table~\ref{t:t1}. The two
  contours in the top and bottom halfs of the density map establish
  the transition between the jet (inner contour: 90\% of jet material)
  and the ambient medium (outer contour: 10\% of jet material). Two Mach discs
  are clearly seen at $z \approx 7.0$ (model JO85) and $z \approx 12.5$
  (model JO86).}
\label{f:JO85-86}
\end{minipage}
\end{center}
\end{figure*}

  Models JO85 and JO86 have an overpressure factor of $K = 5.0$. In
this case, the new equilibrium conditions can not be settled through a 
series of conical shocks and two Mach discs are formed instead. A
snapshot of the two models at $t = 40$ (soon before model JO85 is
completely decollimated and disrupted) is shown in
Fig.~\ref{f:JO85-86}. Besides the difference in the overpressure 
factor between the two sets of models (the relevant one to produce the
switch between conical shocks and Mach discs), models JO85 and JO86
follow a differentially rotating law, with maximum rotation speed
$v_{j, \rm   m}^\phi = 0.34$ (Model JO85) and $0.15$ (JO86). The
effectiveness of Mach discs in decelerating the flow is clearly seen
in the Lorentz factor panel of Fig.~\ref{f:JO85-86}. However,
whereas in the case of model JO86, the flow accross the oblique
magnetosonic shock recollimates downstream and survives at least up to  
$t=100$, model JO85 is completely decollimated and disrupted.

%
%

%
%

\section{Discussion}
\label{s:discussion}

The gist of the paper is the construction of models of planar (i.e., infinite),
axisymmetric, rotating relativistic magnetised jets in transversal
equilibrium made in Sect.~\ref{s:te}. By solving the equation of
transversal equilibrium, we have analyzed the roles of the magnetic
pressure gradient, the magnetic tension and the centrifugal force in
determining the profile of the gas pressure across the jet. Although
our analysis has been performed for a particular choice of radial
dependences of the jet density, velocity and magnetic field, it can be
easily extended to more complex/realistic profiles. However, we should
note that the region of the parameter space explored in our study
leaves outside the class of force-free models \citep[see,
e.g.,][]{Ly99,ML09,ML12}, which in the context of the present study,
should be understood as complementary. 

An interesting result of our analysis is the existence of forbidden
regions in the space of parameters where models of the class
considered in our study could not be settled. These forbidden regions,
represented in the magnetic pitch angle/magnetisation diagram in
Sect.~\ref{s:beta-phi-plane} are associated with the existence of 
maximum axial and toroidal magnetic field components compatible with 
the prescribed equilibrium condition at the jet surface, and/or an
excess of centrifugal force producing gaps with negative pressures in 
the jet. Based on these general grounds, the existence of similar
forbidden regions can be expected for other radial dependencies within 
the class of planar, axisymmetric, magnetised jets in transversal
equilibrium. Whether these forbidden regions can be filled at least
partially with mixed MHD/force-free models (with proper radial
profiles) needs further investigation.

One of the limitations of our approach is the assumption of
the slab symmetry along the jet axis that excludes of our
analysis all the solutions focussing in the collimation and
acceleration of the jet and that have a dependence in the $z$
direction \citep[e.g.,][]{KB07,KV09,Ly10}. Hence our analysis 
on the transversal structure of jets seems better suited for models
beyond the acceleration and collimation region. In this context, the
solutions discussed in this paper can serve as initial injection
conditions for axisymmetric jet simulations. (Magneto)hydrodynamical
simulations of jets have revealed as a very succesful approach to
interpret the varied phenomenology of these objetcs
\citep{GM97,KF97,AG01,AM03,MA09,NG10,PF11,FP12,NM14,MG15}. In
particular, very long baseline interferometric observations of jets often suggest
the presence of quasi-steady features \citep{JM05,LA13}, 
interpreted as recollimation shocks. Besides this, a strong
recollimation shock could be behind the nature of the
millimeter-wavelength radio core in blazars \citep{MJ08,MJ10} in which
the interaction of new superluminal components with 
this shock is invoked to explain the production of $\gamma$-ray flares
in these sources. The same kind of structure although parsecs away
from the radio core has been suggested to explain the
quasi-stationary bright radio feature in the jet of BL Lac located
0.26 mas from the core \citep{CM14,CM15}, the component C80 in
the 3C120 jet \citep{RG10} and the HST-1 complex in M87
\citep{GH12}. 

  The simulations presented in Sect.~\ref{ss:op} show
steady jets with recollimation shocks of different
characteristics. These simulations must be considered as exploratory
since their purpose is only to probe the role of the overpressure
factor and the rotation speed of the jet in the properties of the
internal magnetosonic shocks. According to this, it is important to
note that the parameters used in these simulations were chosen to
produce a variety of internal structures from periodic recollimation
shocks of different strengths and spacing, to isolated Mach discs.
The systematic study of structures like those suggested in the
previous paragraph and their observational signatures with a more
physically oriented selection of parameters is one of the
main objectives of our future work along this line of research.

\section{Summary and conclusions}
\label{s:sc}

  In this paper, we present equilibrium models of relativistic magnetised,
axisymmetric jets with rotation propagating through an homogeneous, 
unmagnetised ambient medium at rest. Under these conditions, the jet
models are characterised by six functions defining the transverse profiles
across the jet radius of the density and the pressure, $\rho(r)$ and
$p(r)$, respectively, and the toroidal and axial  components of the
velocity, $v^\phi(r)$, $v^z(r)$, and of the magnetic field,
$B^\phi(r)$, $B^z(r)$ (the corresponding radial components are
zero by the impossed axisymmetry), and the ambient pressure, $p_a$. In
order to reduce the number of degrees of freedom of our study, we fix
constant the jet rest-mass density and the axial components of the
flow velocity and the magnetic field and analyze the influence of a
toroidal magnetic field and several rotation laws on the structure of
the equilibrium models. The analysis performed in this
  paper can be applied to any radial profiles of density, and axial
  and toroidal flow velocity and magnetic field, in particular to
  those derived from the self-consistently magnetically
  launched RMHD models of \cite{KB07,KV09} or \cite{Ly10}, with
  qualitatively similar conclusions. However these solutions are 
essentially multidimensional and do not belong to the class
of solutions discussed in this work. It is important to note that our
assumptions exclude by construction the important 
  class of force-free equilibrium solutions considered by, e.g.,
  \cite{Ly99,ML09,ML12}, in which the gas pressure and the matter
  inertia are negligible. In
  the context of the present study, the force-free solutions 
  should be understood as complementary. 

  In the case of models without rotation (Sect.~\ref{ss:te-nrj}), the
otherwise constant gas pressure profile is modified by the presence of the
toroidal magnetic field that increases the pressure in a central spine
around the jet axis due to the magnetic tension. The centrifugal force
caused by a rotation velocity (Sect.~\ref{ss:te-rj}) can change this
structure producing a depression inside the jet. Models with large
enough azimuthal speeds and/or small enough azimuthal magnetic fields 
would make the minimum pressure to reach negative values.

  In Sect.~\ref{s:beta-phi-plane}, the set of equilibrium models is
represented in a magnetic pitch angle/magnetisation ($\phi$-$\beta$)
plane in terms of the toroidal and axial magnetic field components,
for fixed jet density and kinematics (axial velocity and rotation
velocity profile) and fixed ambient pressure. Several forbidden
regions are identified in this diagram. First of all, there is a
maximum axial magnetic field compatible with a positive gas pressure
within the jet. Then, for given axial magnetic field, there is also a
maximum toroidal magnetic field compatible with a positive gas pressure at
the jet surface. As a result, there is a forbidden region in the
$\phi$-$\beta$ diagram towards large magnetic pitch angles, large
magnetisations. For rotating jet models there is an additional
forbidden region towards the small pitch angle, small magnetisation
corner in the $\phi$-$\beta$ plane. Models in this region would have
sections with negative gas pressure for given azimuthal velocity profile
due to the centrifugal force. The effect is larger in the case of
models with the maximum azimuthal velocity inside the jet. The present
study can be easily extended to jet models with different transversal
profiles and magnetic field configurations. 

  In Sect.~\ref{s:srmj} we have tested the ability of our RMHD code to
maintain steady equilibrium models of axisymmetric RMHD jets in one and
two spatial dimensions. The one dimensional simulations presented in
Sect.~\ref{ss:pm} can be also understood as a consistency proof of 
the fidelity of the analytical steady solutions discussed in this
paper. Finally, the present study allows us to build
equilibrium jet models with selected properties that could serve as
initial conditions for dynamical (and emission) simulations of
magnetised relativistic jets with rotation, which will be the subject
of future research. In particular, the exploratory simulations presented in
Sect.~\ref{ss:op} were designed to probe the role of the overpressure
factor and the rotation speed of the jet in the properties of the
internal recollimation magnetosonic shocks, invoked to explain several
observational trends in parsec-scale extragalactic jets.

\vspace{5mm}
\noindent
{\it Acknowledgements.}
J.-M. M. acknowledges financial support from the Spanish Ministerio de
Econom\'{\i}a y Competitividad (grants AYA2013-40979-P, and
AYA2013-48226-C3-2-P)  and from the local Autonomous Government
(Generalitat Valenciana, grant Prometeo-II/2014/069). The author also
acknowledges fruitful discussions with profs. J. A. Miralles and
M. Perucho, and the anonymous referee for comments and criticism that
helped to improve this manuscript.

\appendix

\section{RMHD code summary}
\label{a:1}

  Our RMHD code is a conservative, second-order, finite-volume,
constrained-transport code based on high-resolution, shock-capturing
techniques. Its basic ingredients are the following: 

\begin{itemize}

\item[i)] Cell reconstruction: second-order accurate values of
  the primitive variables ${\bf V} = (\rho,p,v^i,B^k)$ at the left and
  right ends of the cells are obtained with linear functions and
  several limiters (MINMOD, VAN LEER, MC). MC and VAN LEER limiters
  can be degraded to MINMOD in case of strong shocks. No jump is
  allowed in the normal component of ${\bf B}$ at a cell boundary and
  the corresponding staggered magnetic field is used.

\item[ii)] Riemann solvers: intercell numerical fluxes are computed by
  means of HLL and HLLC \citep{MB06} Riemann solvers. Accurate bounds
  of the maximum speeds of left and right propagating waves are obtained by
  solving the corresponding characteristic equation for the left and
  right states of each numerical interface.

\item[iii)] Time advance: the multidimensional equations of RMHD are
  advanced in time in an unsplit manner using TVD-preserving
  Runge-Kutta methods of second and third order \citep{SO88,SO89}. The
  time step is determined according to $\displaystyle{\Delta t =
    C\!F\!L \times \min_{i}\left(\frac{\Delta
        r}{|\lambda_{r,i}|}\right)}$ (1D, cylindrical radial version) or to $\displaystyle{\Delta t =
    \frac{C\!F\!L}{\sqrt{2}} \times \min_{i,j}\left(\frac{\Delta
        r}{|\lambda_{r,i,j}|},\frac{\Delta
        z}{|\lambda_{z,i,j}|}\right)}$ (2D, cylindrical axisymmetric
  version), where $\lambda_{r,i,j}$ and $\lambda_{z,i,j}$ are the
  speeds of the fastest waves propagating in cell $i,j$ along the $r$
  and $z$ direction, respectively.

\item[iv)] Constrained transport scheme as in \cite{BS99} for
  multidimensional calculations. The conserved
  variables can be corrected with the algorithms described 
  in \cite{Ma15}.

\item[v)] Primitive variables are recovered as in the 1D$_W$ method of
  \cite{NG06} and solving the resulting equation in $Z = \rho h W^2$
  by bisection.

\end{itemize}

  The code advances the total energy density without the rest-mass 
energy density. This strategy improves the performance
of the conservative scheme when the total energy is dominated by the
rest-mass energy.

  The 1D calculations presented in Sect.~\ref{ss:pm} have
been performed with the MC limiter, and the HLLC Riemann solver,
whereas those shown in Sect.~\ref{ss:op} have been done with the VAN
LEER limiter degraded to MINMOD at shocks, and the HLL Riemann
solver. The advance in time has been done using the third-order
TVD-preserving Runge-Kutta with $C\!F\!L=0.6$ in all the cases. The
relativistic correction algorithm CA2' of \cite{Ma15} has been used to 
correct the conserved variables after each time step in the
2D simulations.

\section{Code validation}
\label{a:2}

  The code passes all the standard RMHD tests in 1D \citep{Ko99a,Ba01,GR06}
and 2D with the expected accuracy and convergence rate. A complete
testing of the code will be presented elsewhere. 

\subsection{Code accuracy}

  The nominal second order of accuracy of the code is verified by means
of the 1D smooth test proposed by \cite{DZ07} describing
the propagation of a large-amplitude, circularly polarized Alfv\'en
wave along a uniform background field ${\bf B}_0$. Taking the
background field along the $x$ axis, the transverse velocity
components are 
\begin{equation}
  v^y = - A \cos [\frac{2\pi}{\lambda} (x -v_a t)], \,\,
  v^z = - A \sin [\frac{2\pi}{\lambda} (x -v_a t)]
\end{equation}
(where $A$ is the amplitude of the wave, and $\lambda$ its wavelength), and 
\begin{equation}
  B^y = - B_0 v^y/v_a, \quad B^z = -B_0 v^z/v_a.
\end{equation}
In the previous expressions, the speed of the Alfv\'en wave, $v_a$, is
given by:
\begin{equation}
  v_a = \pm \sqrt{\frac{B_0^2 (1 - A^2)}{\rho_0 h_0 +B_0^2(1 - A^2)}},
\end{equation}
where $\rho_0$ and $h_0$ are the density and the specific enthalpy of the
background uniform medium\footnote{Note that that expression for $v_a$
  is different from the one in Eq.\,(85) of \cite{DZ07}, though
  equivalent.}.  

\cite{DZ07} utilized this test to assess the order of accuracy of
their code ECHO, and \cite{BS11} did the same for the RMHD
module of ATHENA. We have tested the accuracy of our code by measuring
the errors on one of the transversal velocity components, namely
$v^z$, at one period, $t = \lambda/v_a$, compared to the initial
condition at $t = 0$. Our simulations cover the spatial domain $[0,
2\pi]$, corresponding to a wavelength, with periodic boundary
conditions at both ends. For the background state we choose $\rho_0 =
1$, $h_0 = 5$ ($\gamma = 4/3$), $B_0 = 1$. The amplitude of the
wave is taken as
$$
A = \sqrt{\frac{2}{7 + 3\sqrt{5}}}.
$$
All the values, including the peculiar value of $A$, have been chosen
to define the same wave as the one originally used by \cite{DZ07}.

Table~\ref{t:t2} shows the errors and convergence orders in the $L_1$
norm (the absolute error averaged over the whole computational domain)
for the test at various resolutions with MINMOD, VAN LEER and MC
limiters. The HLLC Riemann solver and the third-order Runge-Kutta with
$C\!F\!L = 0.6$ has been used in all the cases.

%
\begin{table}
\begin{center}
  \caption{Accuracy of the code from the circularly polarized Alfv\'en
    wave test.}
  \begin{tabular}{lrcc}
  \hline
Method & $N$ & $L_1$ error & $L_1$ order \\
 \hline
MINMOD  & $8$        & $1.83 \times 10^{-1}$ & $-$\\
                & $16$      & $8.16 \times 10^{-2}$ & $1.17$\\
                & $32$      & $2.19 \times 10^{-2}$ & $1.90$\\
                & $64$      & $5.47 \times 10^{-3}$ & $2.00$\\
                & $128$    & $1.51 \times 10^{-3}$ & $1.86$\\
                & $256$    & $4.05 \times 10^{-4}$ & $1.90$\\
                & $512$    & $1.05 \times 10^{-4}$ & $1.95$\\
\hline
VAN LEER & $8$        & $1.49 \times 10^{-1}$ & $-$\\
                & $16$      & $3.96 \times 10^{-2}$ & $1.91$\\
                & $32$      & $7.86 \times 10^{-3}$ & $2.33$\\
                & $64$      & $1.58 \times 10^{-3}$ & $2.31$\\
                & $128$    & $3.55 \times 10^{-4}$ & $2.15$\\
                & $256$    & $8.36 \times 10^{-5}$ & $2.09$\\
                & $512$    & $2.01 \times 10^{-5}$ & $2.07$\\
\hline
MC & $8$     & $1.26 \times 10^{-1}$ & $-$\\
      & $16$   & $2.70 \times 10^{-2}$ & $2.22$\\
      & $32$   & $5.45 \times 10^{-3}$ & $2.31$\\
      & $64$   & $1.28 \times 10^{-3}$ & $2.09$\\
      & $128$ & $3.13 \times 10^{-4}$ & $2.03$\\
      & $256$ & $7.75 \times 10^{-5}$ & $2.01$\\
      & $512$ & $1.93 \times 10^{-5}$ & $2.01$\\
\hline
\end{tabular}
\label{t:t2}
\end{center}
\end{table}
%

\subsection{Simulation of discontinuous solutions and thin structures}

\begin{figure*}
  \begin{center}
  \begin{minipage}{176mm}
  \includegraphics[width=16.6cm,angle=0]{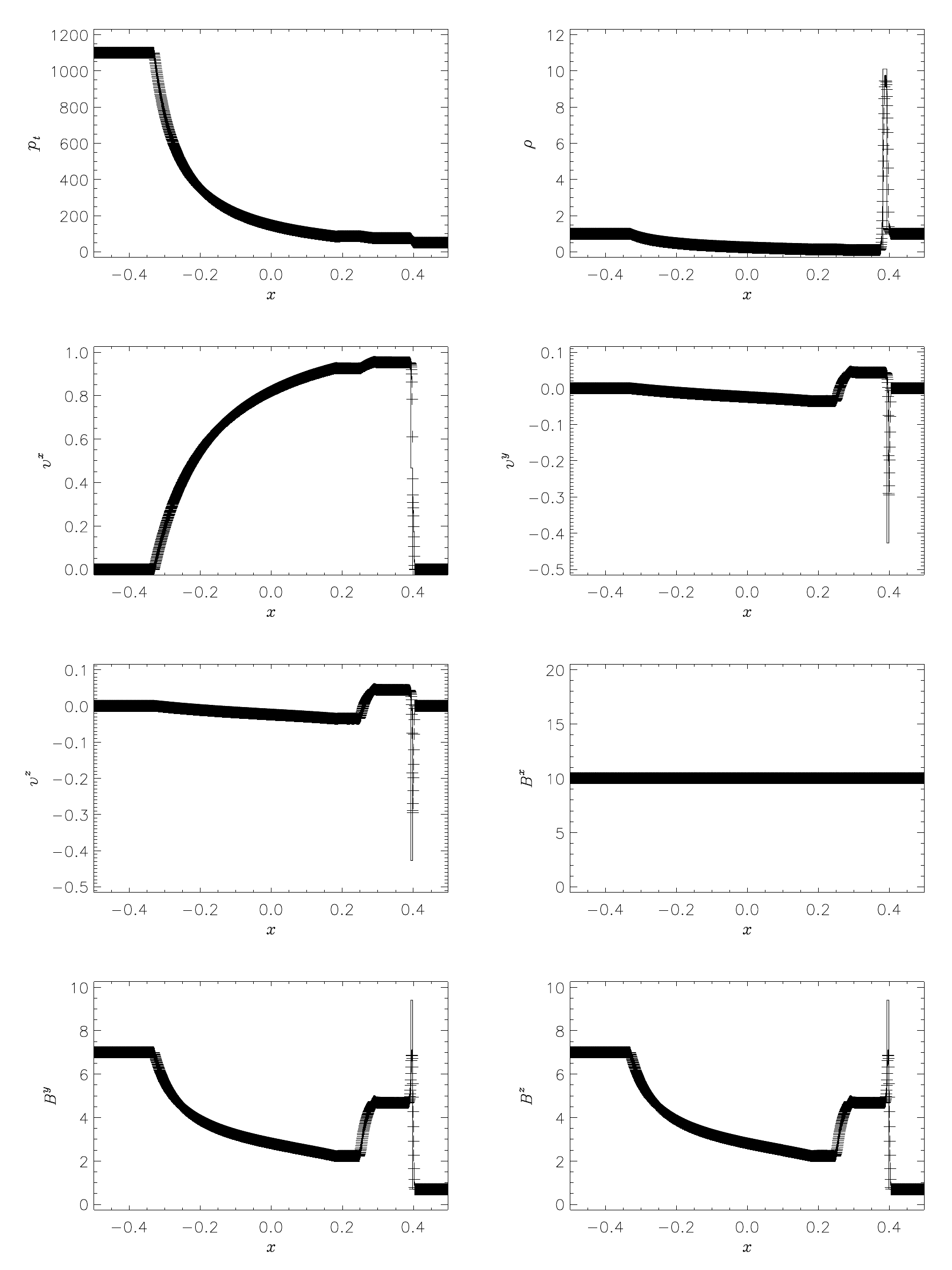}
  \caption{Riemann problem number 3 of Balsara (2001) on 1600 cells with
  the MC limiter, HLLC Riemann solver and Runge-Kutta of third order
  for time advance. The solid line gives the analytical solution as
  computed by Giacomazzo \& Rezzolla (2006).} 
\label{f7}
\end{minipage}
\end{center}
\end{figure*}

The capability of our code in simulating flows involving
discontinuities is demonstrated by means of the Riemann
problem number 3 proposed by \cite{Ba01}, with initial data
$(\rho,p,v^x,v^y,v^z,B^x,B^y,B^z)_L =
(1,1000,0,0,0,10,7,7)$,
$(\rho,p,v^x,v^y,v^z,B^x,B^y,B^z)_R = 
(1,0.1,0,0,0,10,0.7,0.7)$ at both sides of a discontinuity 
at $x=0$. The adiabatic index of the ideal gas equation of state is
$\gamma = 5/3$. According to the analytical solution computed by
\cite{GR06}, the test develops a pair (slow and fast) of left-propagating rarefaction
waves, and a pair of right-propagating shocks, separated by a contact
discontinuity\footnote{As in the rest of coplanar Riemann problems, no 
  Alfv\'en waves develop after the break-up of the initial
  discontinuity.}. A high-density shell forms between the leading 
slow and fast right-propagating shocks (moving at almost the same speed) and the
contact discontinuity. Figure~\ref{f7} shows the numerical results on this
test at $t=0.4$ using 1600 numerical cells in the domain
$x\in[-0.5,0.5]$, with the MC limiter, HLLC Riemann solver and
Runge-Kutta of third order for time advance. The numerical solution is
stable and the high-density shell propagating to the right is captured
quite accurately with the present resolution. The largest errors are
found in the shell values of the transverse components of the flow
velocity and, specially, of the magnetic field. Overall our results
are only improved by those obtained by \cite{MB06}, who used a
second-order, MUSCL-Hancock scheme. 

%
%

\subsection{Multidimensional problems}

Results on the cylindrical magnetised blast wave test \citep{Ko99a}, with
magnetisations larger than $10^8$, can be found in \cite{Ma15}. In
order to test the code in cylindrical coordinates, we have considered
a similar problem simulating the blast of a spherical region embedded
in a uniform, magnetised medium (see \cite{MB07}). In this test, a
sphere of gas with density $\rho = 10^{-2}$ and pressure $p = 1$, is
embedded in a static uniform medium with $\rho = 10^{-4}$ and $p = 3
\times 10^{-5}$. The sphere, with radius $r = 0.8$, is centered at the
origin and a linear smoothing is applied for $0.8 \le r \le 1$. The
whole region is threaded by a constant vertical field in the
$z$-direction, $B_z=1$. The adiabatic index of the ideal gas equation
of state is $\gamma = 4/3$. 

\begin{figure*}
\begin{center}
\begin{minipage}{176mm}
\includegraphics[width=16.6cm,angle=0]{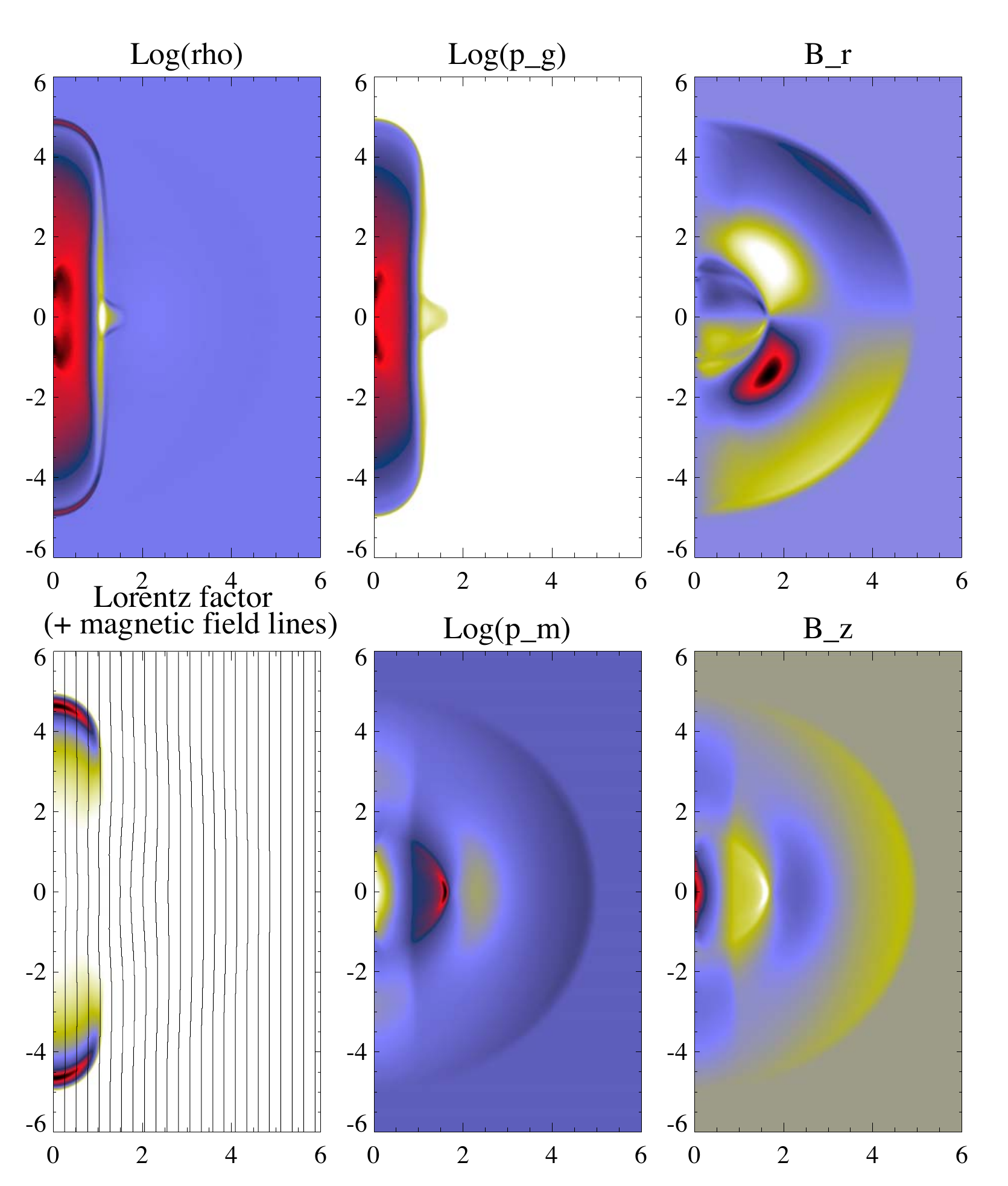}
\caption{Proper rest-mass density, gas pressure and magnetic pressure
  (in logarithmic scale), 
  flow Lorentz factor and magnetic field components at $t=4.0$ for the
  spherical magnetized blast wave test discussed in the text. The
  magnetic field lines are printed on top of the Lorentz factor
  plot. $\log \rho \in [-5.82,-2.58]$, $\log p \in [-4.59,-0.68]$,
  $\log p_m \in [-0.57,-0.16]$, $W \in [1.00,4.57]$, $|B^r| \in [0,
  0.52]$, $|B^z| \in [0.74, 1.19]$.}
\label{f8}
\end{minipage}
\end{center}
\end{figure*}

We have run this test in cylindrical coordinates in the domain $(r,z) \in 
[0,6] \times [-6,6]$ with 512 (radial) $\times$ 1024 (axial)
numerical cells. Open boundary conditions are used along the
boundaries of the computational doain exception made of the axis
($r=0$), were reflecting boundary conditions are used instead. The
same numerical ingredients of the code as for the 2D jet simulations
presented in Sect.~\ref{ss:op} were chosen. Figure~\ref{f8} shows the
distributions of several representative quantities at $t=4.0$. Our
results can be directly compared with those presented by \cite{MB07}
in their Fig.~11. The difference in pressure between the spherical
region and the ambient medium produces the expansion of the central
region delimited by a fast magnetosonic shock propagating radially at
almost the speed of light. Because of the strong sideways magnetic
confinement an elongated structure develops in the $z$ direction with
a maximum Lorentz factor of $W\approx 4.6$. This problem is
particularly challenging because of the very large magnetization
$\beta=1.67 \times 10^4$. The problem also serves to test the
treatment of the geometrical source terms in the code and also the CT
scheme in cylindrical coordinates.

%
%


\label{lastpage}

\end{document}